\documentclass[twocolumn,floatfix,prb,aps,showpacs]{revtex4-2}
\usepackage{graphicx,amsmath,amssymb,color,nicefrac,multirow, makecell, ragged2e,float}
\usepackage[unicode=true,colorlinks=true,linkcolor=blue,citecolor=blue,urlcolor=blue]{hyperref}
\usepackage[titletoc,title]{appendix}

\newcommand{\be}{\begin{equation}}
\newcommand{\ee}{\end{equation}}

\newcommand{\ba}{\begin{eqnarray}}
\newcommand{\ea}{\end{eqnarray}}

\def\beq{\begin{eqnarray}}
\def\eeq{\end{eqnarray}}

\newcommand{\Smat}[1]{S\left(\begin{array}{ccc} #1 \end{array} \right)}

\makeatletter
\newcommand*{\rom}[1]{\expandafter\@slowromancap\romannumeral #1@}
\makeatother

\newcommand{\non}{\nonumber\\}

\usepackage{bm,physics,enumerate,booktabs,ulem}

\newcommand{\imp}{\text{imp}}

\begin{document}
\title{
Disorder-induced topological superconductivity in a spherical
quantum-Hall--superconductor hybrid
}
\author{Koji Kudo, Ryota Nakai, and Kentaro Nomura}
\affiliation{Department of Physics, Kyushu University, Fukuoka 819-0395, Japan}

\begin{abstract}
 Quantum-Hall--Superconductor hybrids 
 have been predicted to exhibit various
 types of topological order, providing possible platforms for intrinsically
 fault-tolerant quantum computing. In this paper, we develop a formulation to 
 construct this hybrid system on a sphere, a useful geometry for identifying 
 topologically ordered states due to its compact and contractible nature.
 As a preliminary step using this framework, 
 we investigate disorder effects on the Rashba-coupled quantum Hall system 
 combined with the type-II superconductor. By diagonalizing the BdG
 Hamiltonian projected into a Rashba-coupled Landau level, we demonstrate the
 emergence of a topological superconducting phase resulting from disorders
 and proximity-induced pairing. Distinctive gapless modes appear in the 
 real-space entanglement spectrum, which is consistent with topological
 superconductivity.
\end{abstract}

\maketitle

\section{Introduction}
The topology of the many-body configuration space determines possible quantum
statistics of particles~\cite{Wu84b}. The fundamental group of the 
configuration space is the symmetry group in three or higher dimensions while 
the braid group in two dimensions, allowing exotic particles beyond bosons and
fermions, namely anyons~\cite{Wilczek82}. The emergence of anyonic 
quasiparticles is a defining feature of topological order~\cite{Wen95}, which 
has revealed a new aspect of phases of matter beyond the scope of Landau's 
theory. Typical examples of topologically ordered states are the fractional
quantum Hall (FQH)
effect~\cite{Tsui82,Laughlin83,Jain07}, quantum spin 
liquids~\cite{Kalmeyer87,Levin05a,Kitaev06,Jackeli09}, and 
$p+ip$ superconductors (SCs)~\cite{Read00,Ivanov01}.
Recent progress in experiments for anyons has been made, for example, through 
measurements of the half-integer quantized thermal Hall conductivity in the 
Kitaev materials~\cite{Kasahara18,Yokoi21,Bruin22} and detections of the fractional statistics via directly braiding the FQH 
quasiparticles~\cite{Nakamura20,Bartolomei20,Nakamura23}. Creating, 
manipulating, and reading non-Abelian anyons are basic elements for topological
quantum computing~\cite{Kitaev03,Nayak08,Jain20b}, which has been an ultimate 
goal in condensed matter physics.   

A key strategy for generating non-Abelian anyon platforms has involved 
designing topological materials. Hybridizing well-understood ingredients, even 
if they are not inherently topological, has proved to be a 
valuable tool in this process, leveraging their interplay to 
introduce topological order. 
For instance, hybridization with $s$-wave non-topological SCs induces 
superconducting proximity effects, leading to Majorana 
modes on surfaces of strong topological insulators~\cite{Fu08,Akhmerov09}, 
spin-orbit coupled semiconductors~\cite{Sau10,Lutchyn10,Oreg10,Alicea10}, quantum 
anomalous Hall systems~\cite{Qi10}, and integer quantum Hall (IQH)
systems~\cite{Bjorn16,Mishmash19,Chaudhary20}. 
Recent efforts have been
directed towards exploring hybrid quantum-Hall--SC (QH-SC) systems that host 
much more exotic particles such as parafermions~\cite{Cheng12,Lindner12,Burrello13,Clarke13,Vaezi13,Milsted14,Klinovaja14,Alicea16,Sagi17,Liang19} and 
Fibonacci anyons~\cite{Vaezi14,Mong15,Hu18,Lopes19,Gul22}
(The so-called QH 
superconductivity~\cite{
Teifmmode89,Rajagopal91,Teifmmode91,Norman91,Akera91,Rajagopal91b,Rasolt92,MacDonald92,Norman92,Rajagopal92,MacDonald93,Ryan93,Ryan93b,Norman95,Maifmmode02,Scherpelz13,Ran19,Kim19,Chaudhary21,Schirmer22,Schirmer22a}
also exhibits similar physics as the QH-SC hybrids).
These developments have highlighted the potential of QH-SC hybrids for 
universal topological quantum computation.

In theoretical exploration of topological order, the geometry of systems is
quite crucial. Particularly, boundaries introduce low-energy modes as 
edge states~\cite{Wen90a,Hatsugai93}, rendering compact surfaces suitable for 
investigating bulk properties of topologically ordered states. On a compact 
surface with genus $g\geq1$, the one-dimensional unitary representation of the 
braid group is absent. Consequently, states on such surfaces must be degenerate
to form a multi-component 
structure~\cite{Birman69,Einarsson90,Wen90,Oshikawa06,Sato06,Oshikawa07}, 
referred to as topological degeneracy~\cite{Haldane85b,Wen90d}.
This phenomenon, while intriguing and closely related to topological nature 
of anyons, can pose technical challenges in numerical studies. 
Indeed, the spherical geometry~\cite{Haldane83} with $g=0$ has been commonly 
used in the FQH physics.

\begin{figure}[b]
\includegraphics[width=\columnwidth]{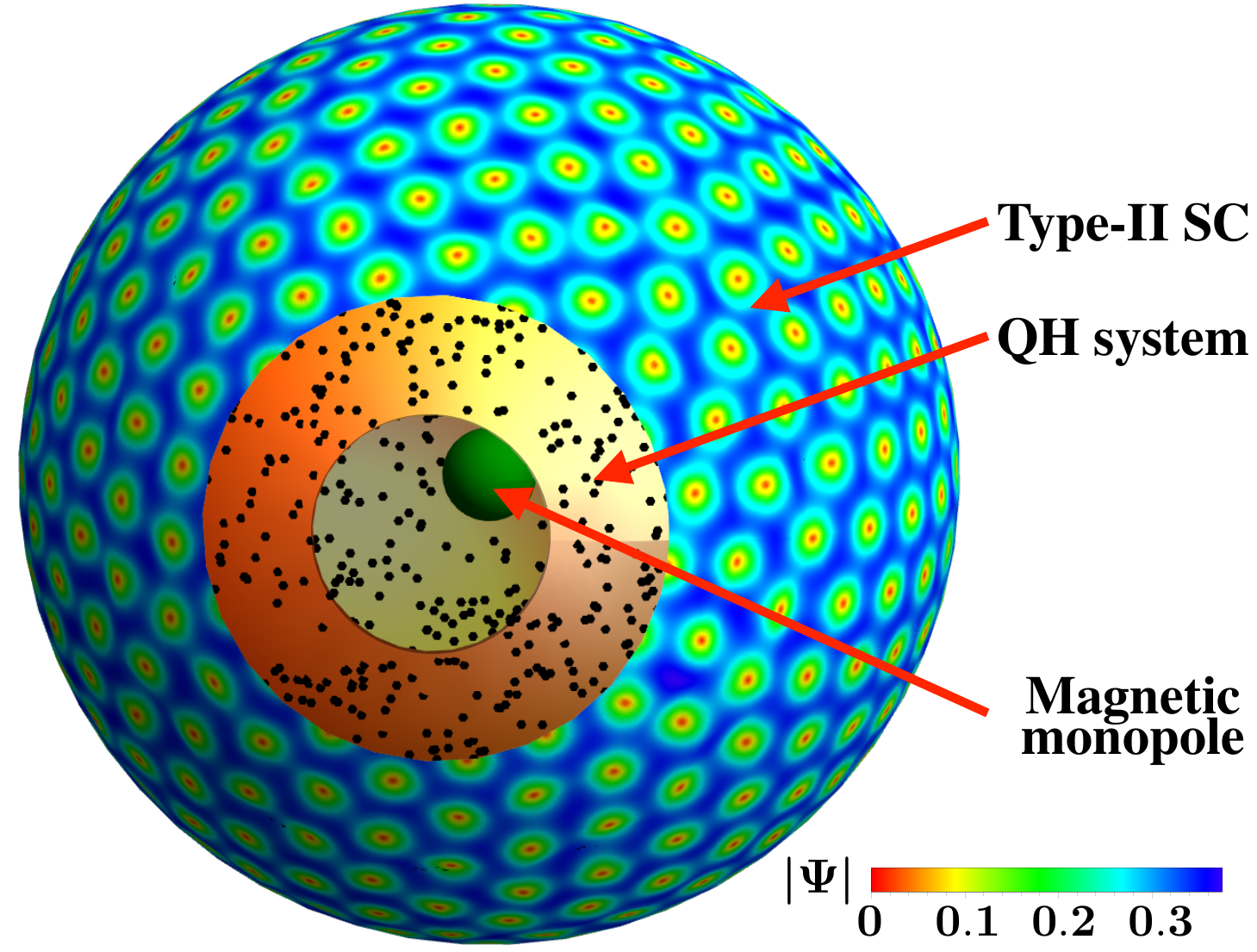}
 \caption{QH--SC hybrid on a sphere. The outer (inner) surface represents the 
 type-II SC (QH system). The color of the outer surface indicates the order parameter $|\Psi|$ 
 at $2\bar{Q}=450$. We normalize $\Psi$ as $\int d\vec{\Omega}|\Psi|^2=1$ for 
 simplicity. The black dots in the inner surface depict impurities, with a 
 total count of 4500. The 
 order (outer or inner) of each system is not significant.  The green sphere at
 the center is the magnetic monopole.
 }
 \label{fig:hybrid}
\end{figure}
This background motivates us to develop a formulation to construct a hybrid
QH-SC system with the spherical geometry. As a preliminary step, we consider
the Rashba-coupled IQH system combined with the type-II SCs, motivated by
Refs.~\onlinecite{Mishmash19,Chaudhary20}. In addition, we
incorporate 
random distributions of $\delta$-function 
impurities. Figure~\ref{fig:hybrid} visually represents our model. The type-II 
SC is chosen because of the strong magnetic field required for the QH system. 
Our 
numerical studies demonstrate that the interplay between disorders and 
proximity-induced pairing results in a topological 
superconducting phase associated with the half-integer Chern number
[equivalently the unit Bogoliubov--de Gennes (BdG) Chern number]. We 
identify this phase by detecting gap-closing lines. The entanglement spectrum 
reveals distinctive gapless modes, providing further evidence for 
topological superconductivity.

The paper is organized as follows. In Sec.~\ref{sec:Abrikosov}, we review two
components in our hybrid system on a sphere: the Rashba-coupled QH system and a
model type-II SC. In Sec.~\ref{sec:model}, we construct the hybird QH-SC 
system. Section~\ref{sec:results} presents numerical results,
and the paper concludes in Sec.~\ref{sec:conc}.

\section{Setup for the spherical geometry}
\label{sec:Abrikosov}
\subsection{Rashba-coupled Landau level}
We begin by considering the Rashba-coupled Landau level (LL) structure. The single-particle 
Hamiltonian reads 
$H_1=\vec{\pi}^2/2m-\alpha_R\left(\vec{\sigma}\times\vec{\pi}\right)$,
where $\vec{\pi}$ is the canonical momentum, $\vec{\sigma}$ is the Pauli
matrices, and $\alpha_R$ is the Rashba coupling strength. In the planar 
geometry, this reduces 
to~\cite{Rashba60,Yu84,Schliemann03,Shen04,Ito12,Mishmash19,Chaudhary20}
\begin{align}
 H_1
 &=\hbar\omega_c\left(                                                   
 \begin{array}{cc}                                                       
  a^\dagger a+\frac{1}{2} & -ig_Ra \\
  ig_Ra^\dagger & a^\dagger a+\frac{1}{2}
 \end{array}
 \right),
 \label{eq:ham1}
\end{align}
where $\omega_c=eB/mc$ is the cyclotron frequency and 
$g_R={\sqrt{2}\alpha_R}/{l_B\omega_c}$ with $l_B$ the magnetic strength.
The ladder operator is defined by 
$a^\dagger=\left(\pi_x+i\pi_y\right)l_B/\sqrt{2}\hbar$. Within the subspace,
\begin{align}
 \Phi_{nm}=\left(\ket{n-1,m},\ket{n,m}\right)
 \text{ with $n\geq1$},
 \label{eq:basis}
\end{align}
where $\ket{n,m}$ is the eigenstate of angular momentum 
$\hbar m$ in the $n$th LL without spin-orbit coupling, the 
Hamiltonian $H_1$ is block-diagonalized as
\begin{align}
 \Phi_{nm}^\dagger H_1\Phi_{nm}
 &=\hbar\omega_c\left(
 \begin{array}{cc}
  n-\frac{1}{2} & -ig_R\sqrt{n} \\
  ig_R\sqrt{n} & n+\frac{1}{2}
 \end{array}
 \right).
 \label{eq:Esys}
\end{align}
Its eigenvalues and eigenvectors are
\begin{align}
 \begin{split}
  &\epsilon_{n\pm}=\hbar\omega_c\left(n\pm\sqrt{1/4+ g_R^2n}\right),\\
  &\vec{v}_{n\pm}=
  \left(
  \begin{array}{c}
   i/2\mp i\sqrt{g_R^2n+1/4} \\
   g_R\sqrt{n}
  \end{array}                                                             
  \right)/\mathcal{N},
 \end{split}
 \label{eq:ev}
\end{align}
where $\mathcal{N}$ is a normalization factor. 
In addition, the unpaired state $\left(0,\ket{0,m}\right)$ is an eigenstate
of $H_1$ with energy $\epsilon_0=\hbar\omega_c/2$.
In the limit $g_R\rightarrow\infty$, $H_1$ reduces to the Hamiltonian of 
massless Dirac fermions. 

\begin{figure}
 \includegraphics[width=\columnwidth]{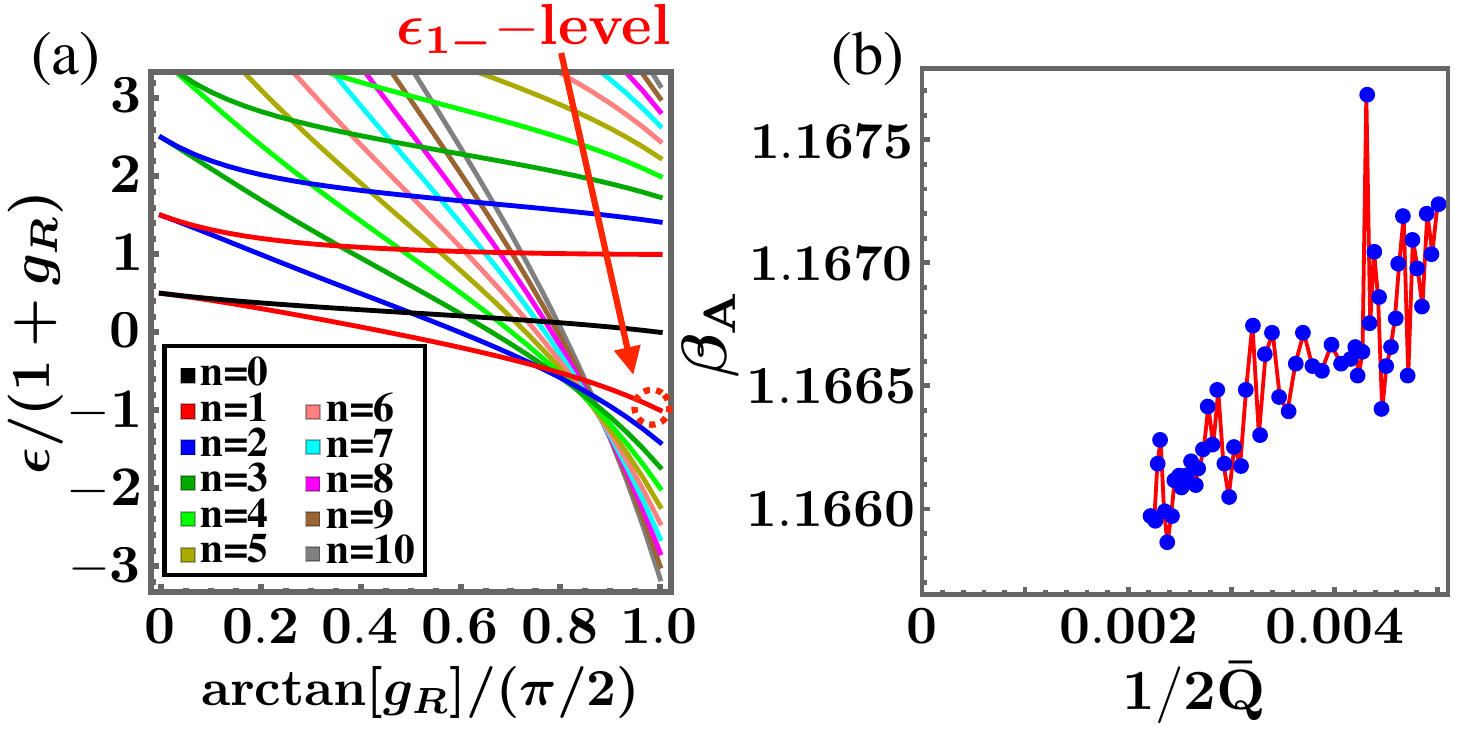}
 \caption{
 (a) Rashba-coupled LLs $\epsilon_0$ and $\epsilon_{n\pm}$ with $n\leq10$.
 (b) Abrikosov factor $\beta_A$ at various $2\bar{Q}$'s.
 }
 \label{fig:Rashba-betaA}
\end{figure}
Figure~\ref{fig:Rashba-betaA}(a) shows the single-particle energy. To observe 
the evolution for
$g_R\in[0,\infty)$, we scale the energy by $1+g_R$ and plot it as a function of
$\arctan(g_R)$. 
At $g_R\sim0$, the energy is quantized in increments of $n$ while 
in increments of $\sqrt{n}$ at $g_R\sim\infty$. In the next section, 
we will focus on the Rashba-coupled LL with $\epsilon_{1-}$, 
referred to as $\epsilon_{1-}$-level.
This is the lowest energy level for 
$0<g_R<\sqrt{6}$ and becomes the ``$n=-1$ LL'' of the two-dimensional
Dirac Hamiltonian~\cite{Neto09}
as $g_R$ approaches infinity.

In the following, we consider Haldane's spherical geometry~\cite{Haldane83}, 
where $N$ particles move on the surface under a radial magnetic field. The 
total 
radial flux is $2Q\phi_0$, where $\phi_0=hc/e$ is the flux quantum and $2Q$ is 
an integer. In the spinless problem without spin-orbit coupling, 
single-particle 
states are labeled by the orbital angular momentum $l$ and its $z$-component 
$m$ because of rotational symmetry. Their possible values are 
$l=|Q|,|Q|+1,\ldots$ and $m=-l,-l+1,\ldots,l$. The $2l+1$ states with 
$l=|Q|+n$ corresponds to the $n$th LL. The eigenstates are the
monopole harmonics $Y_{Qlm}(\vec{\Omega})$~\cite{Wu76,Wu77}, where 
$\vec{\Omega}$ represents the angular coordinates $\theta$ and $\phi$.


Spin-orbit coupling is not straightforward to apply as it mixes different LLs having different degrees of degeneracy on a sphere.
The same issue also arises, e.g. in the QH physics in graphene~\cite{Nomura06,Goerbig06,Apalkov06,Toke06,Toke07,Balram15c}.
Fortunately, because of the non-flat geometry, relativistic electrons on a Haldane's sphere with physical fluxes $Q$ are subject to different magnetic fluxes $Q_\pm=Q\pm 1/2$ depending on the spin orientation~\cite{Jellal08,Arciniaga16,Hasebe16,Yonaga16}.
As a result, $n$th LL with $Q_+$ and $(n+1)$th LL with $Q_-$ have the same degree of degeneracy.
Based on this solution, we use the following basis, instead of 
Eq.~\eqref{eq:basis}, for the spherical geometry:
\begin{align}
 \Phi_{Qnm}=\left(Y_{Q_{+},Q_{+}+n,m},Y_{Q_{-},Q_{+}+n,m}\right).
\end{align}
The labels $n,m$ represent the Landau index and the angular momentum.

\subsection{Abrikosov vortex lattice}
We now review a model type-II superconductor on a 
sphere~\cite{Dodgson_96,Dodgson97} to calculate the superconducting 
order parameter, which we will use to construct the proximity-induced pairing 
amplitude in our hybrid system below. Here we consider a clean system with a 
magnetic field slightly smaller than the upper critical field $H_{c2}$.
A magnetic monopole is placed at the center of the sphere in the same fashion 
as above. For clarity, we mark quantities for Cooper pairs by a bar; 
superconducting flux quantum $\bar{\phi}_0=hc/\bar{e}$ with $\bar{e}=2e$ and 
the total flux $2\bar{Q}\bar{\phi}_0$ (equivalently, $\bar{Q}=2Q$).

To identify the superconducting order parameter, we consider the GL free energy
$\mathcal{F}\left[\Psi\right]=\int d\vec{\Omega}\,f\left[\Psi\right]$ with
\begin{align}
f\left[\Psi\right]
 =
 -a|\Psi|^2
 +\frac{b}{2}|\Psi|^4
 +\frac{1}{2m}
 \left|\left(\frac{\hbar}{i}\vec{\nabla}
 +\frac{\bar{e}}{c}\vec{A}\right)\Psi\right|^2
 +\frac{|\vec{B}|^2}{8\pi},
 \label{eq:GLF}
\end{align}
where $a,b$ are phenomenological parameters ($a,b>0$). We now demonstrate that
$\Psi$ minimizing
$\mathcal{F}$ does not depend on $a$ and 
$b$ except the overall amplitude. 
Close to the upper critical field,
the order parameter lies in the
lowest LL of Cooper pairs as
\begin{align}
 \Psi(\vec{\Omega})
 =\sum_{\bar{m}=\bar{Q}}^{\bar{Q}}u_{\bar{m}}
 Y_{\bar{Q}\bar{Q}\bar{m}}(\vec{\Omega}),
\end{align}
and one obtains
\begin{align}
 -a\int d\vec{\Omega}|\Psi|^2
 &=-a\vec{u}^\dagger\vec{u},\\
 \frac{b}{2}\int d\vec{\Omega}|\Psi|^4
 &=\frac{b}{2}\sum_{s=-2\bar{Q}}^{2\bar{Q}}\left|
 \vec{u}^TK^{(s)}\vec{u}\right|^2,
\end{align}
where $\vec{u}=\left(u_{-\bar{Q}},u_{-\bar{Q}+1},\ldots,u_{\bar{Q}}\right)$
and,
\begin{align}
 &K^{(s)}_{ij}
 =\Smat{\bar{Q}&\bar{Q}&-2\bar{Q} \\ \bar{Q}&\bar{Q}&2\bar{Q} \\i&j&-i-j}
 \delta_{s,i+j},\\
 &\Smat{Q_1&Q_2&Q_3 \\ l_1&l_2&l_3 \\m_1&m_2&m_3}       
 \equiv
 \int d\vec{\Omega}
 Y_{Q_1l_1m_1}Y_{Q_2l_2m_2}Y_{Q_3l_3m_3}.
\end{align}
Dropping the constant, the GL free energy reduces to
\begin{align}
 \mathcal{F}(\vec{u})=-ax+\frac{b}{2}\frac{\beta_A}{4\pi}x^2,
 \label{eq:finalF}
\end{align}
where $x=\vec{u}^\dagger\vec{u}$ and $\beta_A$ is the Abrikosov factor:
\begin{align}
 \beta_A\equiv\frac{\langle|\Psi|^4\rangle}{\langle|\Psi|^2\rangle^2}
 =\frac{4\pi\sum_s\left|\vec{u}^TK^{(s)}\vec{u}\right|^2}
 {\left(\vec{u}^\dagger\vec{u}\right)^2}.
 \label{eq:betaA}
\end{align}
Here $\langle \cdot\rangle$ represents a spatial average. 
Equation~\eqref{eq:betaA} implies that $\beta_A$ is independent of
$|\vec{u}|$ and also $x$ ($=\vec{u}^\dagger\vec{u}$). The GL free energy has a 
peak at $x=4\pi a/(b\beta_A)\equiv x_0$ and its value is 
\begin{align}
 \mathcal{F}_\text{peak}(\vec{u})=-\frac{2\pi a^2}{b\beta_A}.
\end{align}
Thus, minimizing $\mathcal{F}$ is equivalent to minimizing 
$\beta_A$. Since $\beta_A$ does not involve $a$ and $b$,
the solution of $\vec{u}$ that minimizes $\mathcal{F}$ does not either, apart 
from the norm (the norm of $\vec{u}$ is determined by $x_0$.)

We numerically minimize $\mathcal{F}$ with $a=b=1$ and calculate the Abrikosov
factor $\beta_A$ for various $2\bar{Q}$ in Fig.~\ref{fig:Rashba-betaA}(b). The 
value 
of $\beta_A$ with
$2\bar{Q}\rightarrow\infty$ is $1.1652$, determined through linear 
extrapolation.
This is 
slightly larger than $1.1596$ calculated on an infinite flat 
plane with triangular Abrikosov vortex lattices~\cite{Kleiner64}. This 
deviation comes from the fact that a triangular lattice cannot generally cover
a surface of a sphere~\cite{Dodgson_96}. In Fig.~\ref{fig:hybrid}, we 
plot $|\Psi(\vec{\Omega})|$ on a sphere at
$2\bar{Q}=450$. The vortices basically form a triangular lattice but there are
defects as well. Similar discussions can be found in the context of the famous 
Thomson problem~\cite{Thomson04,Wales06,Wales09} and the Wigner crystal in the 
QH problem~\cite{Zhao18}.

We use the solution of $\vec{u}$ to construct proximity-induced pairing 
amplitude in our hybrid system below. In Fig.~\ref{fig:Rashba-betaA}(b),
we minimize $\mathcal{F}$ 
with $M_{\text{sample}}$ different initial points of $\vec{u}$
with $80\leq M_{\text{sample}}\lesssim400$. We then plot $\beta_A$ if the lowest $\beta_A$ is 
at least 3-fold degenerate. Here, two $\beta_A$'s are considered the same if 
their difference is less than $10^{-8}$. The magnetic fluxes are set as
$\bar{Q}=100,101,102,\ldots,119$ and $120,123,126,\ldots,225$. (We
failed to seek $\beta_A$ that satisfies this criterion at 
some values of $\bar{Q}$'s.) 

\section{Spherical Hybrid system}
\label{sec:model}
Our spherical hybrid system is composed of a Rashba-coupled QH system with 
disorders and a type-II $s$-wave SC. Hybridization induces the superconducting 
proximity effect on the QH system. We write its total Hamiltonian as
\begin{align}
 \mathcal{H}
 &=\int d\vec{\Omega}\,
 c^\dagger(\vec{\Omega})\left[
 H_1(\vec{\Omega})+H_\imp(\vec{\Omega})-\mu
 \right]c(\vec{\Omega})\non
 &\qquad\qquad\qquad+\int d\vec{\Omega}\,
 c_\uparrow^\dagger(\vec{\Omega})\Delta(\vec{\Omega})
 c_\downarrow^\dagger(\vec{\Omega})+\text{h.c.},
 \label{eq:originalH}
\end{align}
where $c^\dagger_\sigma(\vec{\Omega})$ is the creation operator for a fermion
with spin $\sigma$, $c^\dagger=(c^\dagger_\uparrow,c^\dagger_\downarrow)$,
$\mu$ is a chemical potential, and $\Delta(\vec{\Omega})$ is the pairing 
amplitude. Here, $H_\imp$ represents a random distribution of $2N_\imp=20Q$ 
impurities at positions $\vec{\Omega}_i$'s with the energy $\pm w$:
\begin{align}
 H_\imp(\vec{\Omega})
 &=\sum_{i=1}^{2N_\imp}(-1)^iw\delta^2(\vec{\Omega}-\vec{\Omega}_i).
\end{align}
The impurities broaden the LLs and their width is estimated to be 
$\Gamma=(w/R^2)\sqrt{4\rho/2\pi l_B^2}$, where $R=l_B\sqrt{Q}$ and 
$\rho=2N_\imp/4\pi R^2$, by using the self-consistent Born 
approximation~\cite{Ando83}. We parameterize the strength
of disorders by $\Gamma$ rather than $w$ below.
The pairing amplitude $\Delta(\vec{\Omega})$ in Eq.~\eqref{eq:originalH} is 
defined by
\begin{align}
 \Delta(\vec{\Omega})
 &=C\sum_{\bar{m}=-\bar{Q}}^{\bar{Q}}u_{\bar{m}}
 Y_{\bar{Q}\bar{Q}\bar{m}}(\vec{\Omega}),
\end{align}
where $C$ is a constant and $u_{\bar{m}}$ is the solution of the 
minimizing problem of the GL free energy $\mathcal{F}$ discussed above. 
We parameterize the strength of the proximity effect by 
$\Delta_0=\sqrt{|C|^2\vec{u}^\dagger\vec{u}/(4\pi)}$, which
corresponds to the spatial average of $\Delta(\vec{\Omega})$, i.e.,
\begin{align}
 \int d\vec{\Omega}\left|\Delta(\vec{\Omega})\right|^2=4\pi \Delta_0^2.
\end{align}

Hereafter, we consider the BdG Hamiltonian projected into the 
$\epsilon_{1-}$-level as
\begin{align}
 &\mathcal{H}_\text{BdG}
 =\frac{1}{2}\bm{f}^\dagger
 H_\text{BdG}
 \bm{f},
 \label{eq:BdG}\\
 &H_\text{BdG} = \left(
 \begin{array}{cc}
  h_0 & 2D \\
  2D^\dagger & -h_0^*
 \end{array}
 \right),
\end{align}
where $h_0$ and $D$ are $(2Q_++1)\times(2Q_++1)$ matrices,
$\bm{f}^\dagger
=\left(f^\dagger_{-Q_+},\ldots,f^\dagger_{Q_+},f_{-Q_+},\ldots,f_{Q_+}\right)$,
and $f^\dagger_m$ is the creation operator of states in $\epsilon_{1-}$-level
defined by
\begin{align}
 &f^\dagger_m
 =\left(d^\dagger_{Q_+Q_+m\uparrow},d^\dagger_{Q_-Q_+m\downarrow}\right)
 \vec{v}_{1-},\\
 &d^\dagger_{Q_\pm lm\sigma}=\int d\vec{\Omega}Y_{Q_\pm lm}(\vec{\Omega})
 c_{\sigma}^\dagger(\vec{\Omega}).
 \label{eq:d}
\end{align}
Here, $Q_\pm$ in Eq.~\eqref{eq:d} takes $Q_+$ ($Q_-$) with spin 
$\sigma=\uparrow(\downarrow)$.
This projection is based on two assumptions:
first, the chemical potential $\mu$ is close to $\epsilon_{1-}$-level, and 
second, the Landau gap $\hbar\omega_c$ significantly exceeds both disorder and 
proximity effect. 
Let us now calculate the matrices $h_0$ and $D$, and discuss some properites of
the BdG spectrum.


\subsection{matrix $h_0$}
The matrix $h_0$ is derived from the first term of 
$\mathcal{H}$ in Eq.~\eqref{eq:originalH}. 
Within the $\epsilon_{1-}$-level space, we can perform 
the following replacement,
\begin{align}
 &
 c^\dagger_\sigma(\vec{\Omega})
 \rightarrow\sum_{m=l}^{l}Y_{Q_\pm lm}^*(\vec{\Omega})
 d_{Q_\pm lm\sigma}^\dagger,\\
 &\left(d^\dagger_{Q_+Q_+m\uparrow},d^\dagger_{Q_-Q_+m\downarrow}\right)
 \rightarrow f^\dagger_m\vec{v}_{1-}^\dagger.
\end{align}
Then the disorder potential is expressed by 
\begin{align}
 \int d\vec{\Omega}\,
 c^\dagger(\vec{\Omega})H_\imp(\vec{\Omega})c(\vec{\Omega})
 =\sum_{m,m'=-Q_+}^{Q_+}W_{mm'}f_m^\dagger f_m,
\end{align}
where 
\begin{align}
 \begin{split}
  &W_{mm'}
  =\left|\left[\vec{v}_{1-}\right]_1\right|^2
  W_{Q_+}(Q_+m;Q_+m')\\
  &\qquad\qquad\qquad\qquad
  +\left|\left[\vec{v}_{1-}\right]_2\right|^2W_{Q_-}(Q_+m;Q_+m'),\\
  &W_{Q}(lm;l'm')
  =w(-1)^{Q-m'}
  \sum_{l''=l_i}^{l_f}\times\\
  &\ \ 
  \Smat{-Q&Q&0 \\ l&l'&l'' \\ -m&m'&m-m'}
  \sum_{i=1}^{2N_\text{imp}}
  (-1)^i
  Y_{l''m'-m}(\vec{\Omega}_i),
 \end{split}
\end{align}
where $l_i=\max\{|l-l'|,|m-m'|\}$ and $l_f=l+l'$. Noting that $H_1$ becomes an
identity matrix with a prefactor $\epsilon_{1-}$, one gets
\begin{align}
 \left(h_0\right)_{mm'}=W_{mm'}+\delta_{mm'}(\epsilon_{1-}-\mu).
\end{align}

\subsection{matrix $D$}
The pairing amplitude is deformed as
\begin{align}
 &\int d\vec{\Omega}\,
 c_\uparrow^\dagger(\vec{\Omega})\Delta(\vec{\Omega})
 c_\downarrow^\dagger(\vec{\Omega})
 =\sum_{ll'}\sum_{mm'}\Delta(Q_+lm;Q_-l'm')\times\non
 &\qquad\qquad\qquad\qquad\qquad\qquad\qquad
 d^\dagger_{Q_+lm\uparrow}d^\dagger_{Q_-l'm'\downarrow}
\end{align}
where
\begin{align}
 &\Delta(Q_+lm;Q_-l'm')
 =\int d\vec{\Omega}Y^*_{Q_+lm}Y^*_{Q_-l'm'}\Delta(\vec{\Omega})\non
 &
 =Cu_{m+m'}(-1)^{\bar{Q}-m-m'}
 \Smat{-Q_+&-Q_-&\bar{Q} \\ l&l'&\bar{Q}\\-m&-m'&m+m'}.
\end{align}
Noting the validity of the replacement 
$d^\dagger_{Q_+lm\uparrow}d^\dagger_{Q_-l'm'\downarrow}
\rightarrow i\gamma f_m^\dagger f_{m'}^\dagger$ within the 
$\epsilon_{1-}$-level, where
$i\gamma
\equiv\left[\vec{v_{1-}}\right]_1^*\left[\vec{v_{1-}}\right]_2^*
=-g_R/\sqrt{1+4g_R^2}$, one gets
\begin{align}
 D_{mm'}
 =i\gamma
 \Delta(Q_+Q_+m;Q_-Q_+m').
\end{align}
One can easily show $D^T=-D$.
The parameter $\gamma$ is a monotonically decreasing function of
$g_R$, with $\gamma(0)=0$ and $\gamma(\infty)=-1/2$. This implies that the
strength of spin-orbit coupling enhances the proximity-induced superconducting 
pairing~\cite{Mishmash19,Chaudhary20}. For the purpose of demonstrating topological superconductivity, we 
fix $g_R=10^{10}$ in the numerical calculations below.
In this regime, the Rashba-coupled system can be replaced with monolayer 
graphene or a surface of a three-dimensional topological insulator with 
magnetic fields.

\subsection{BdG spectrum}
The BdG spectrum is given by 
\begin{align}
 H_\text{BdG}
 \left(
 \begin{array}{c}
  u_k \\
  v_k
 \end{array}
 \right)
 =E_k
  \left(
 \begin{array}{c}
  u_k \\
  v_k
 \end{array}
 \right),
\end{align}
where $u_k$ and $v_k$ are $N$-dimensional vectors and $2N\equiv 4Q_++2$ is the 
dimension of $H_\text{BdG}$.
Here, we arrange the eigenvalues in ascending order as
$E_{-(N-1/2)}\leq E_{-(N-1/2)+1}\leq\ldots\leq E_{N-1/2}$. The particle-hole 
symmetry brings $E_{-k}=-E_{k}$ and 
$
\left(
\begin{array}{c}
 u_{-k} \\
 v_{-k}
\end{array}
\right)
=\left(
\begin{array}{c}
 v_k^* \\
 u_k^*
\end{array}
\right)
$.
Then, we have
\begin{align}
 \mathcal{H}_\text{BdG}
 =&\frac{1}{2}\sum_{\text{all }k}
 E_kg_k^\dagger g_k
 =\sum_{k<0}E_kg_k^\dagger g_k+\text{const},
\end{align}
where
\begin{align}
 g_{k}^\dagger
 =&g_{-k}
 =\bm{f}^\dagger
 \left(
 \begin{array}{c}
  u_k \\
  v_k
 \end{array}
 \right).
\end{align}
The ground state is given by $\ket{G}=\prod_{k<0}g^\dagger_k\ket{0}$ with
$\ket{0}$ the vacuum state of electrons. We will use the following matrices,
\begin{align}
 \begin{split}
  &u\equiv\left(u_{-(N-1/2)},\ldots,u_{-1/2}\right),\\
  &v\equiv\left(v_{-(N-1/2)},\ldots,v_{-1/2}\right),
 \end{split}
 \label{eq:uv}
\end{align}
to calculate the entanglement spectrum below.

\section{Numerical results}
\label{sec:results}
The system parameters we have not fixed yet are
\begin{align}
 (2Q,\Delta_0,\Gamma,\mu).\nonumber
\end{align}
The main goal now is to find a topological superconducting phase by varying 
$\Delta_0/\Gamma$ and $\mu/\Gamma$. For simplicity, we fix $\Gamma=1$ and 
measure $\mu$ relative to $\epsilon_{1-}$. The numerical results below are all 
obtained by diagonalizing $\mathcal{H}_\text{BdG}$ in Eq.~\eqref{eq:BdG}.

\begin{figure}
 \includegraphics[width=\columnwidth]{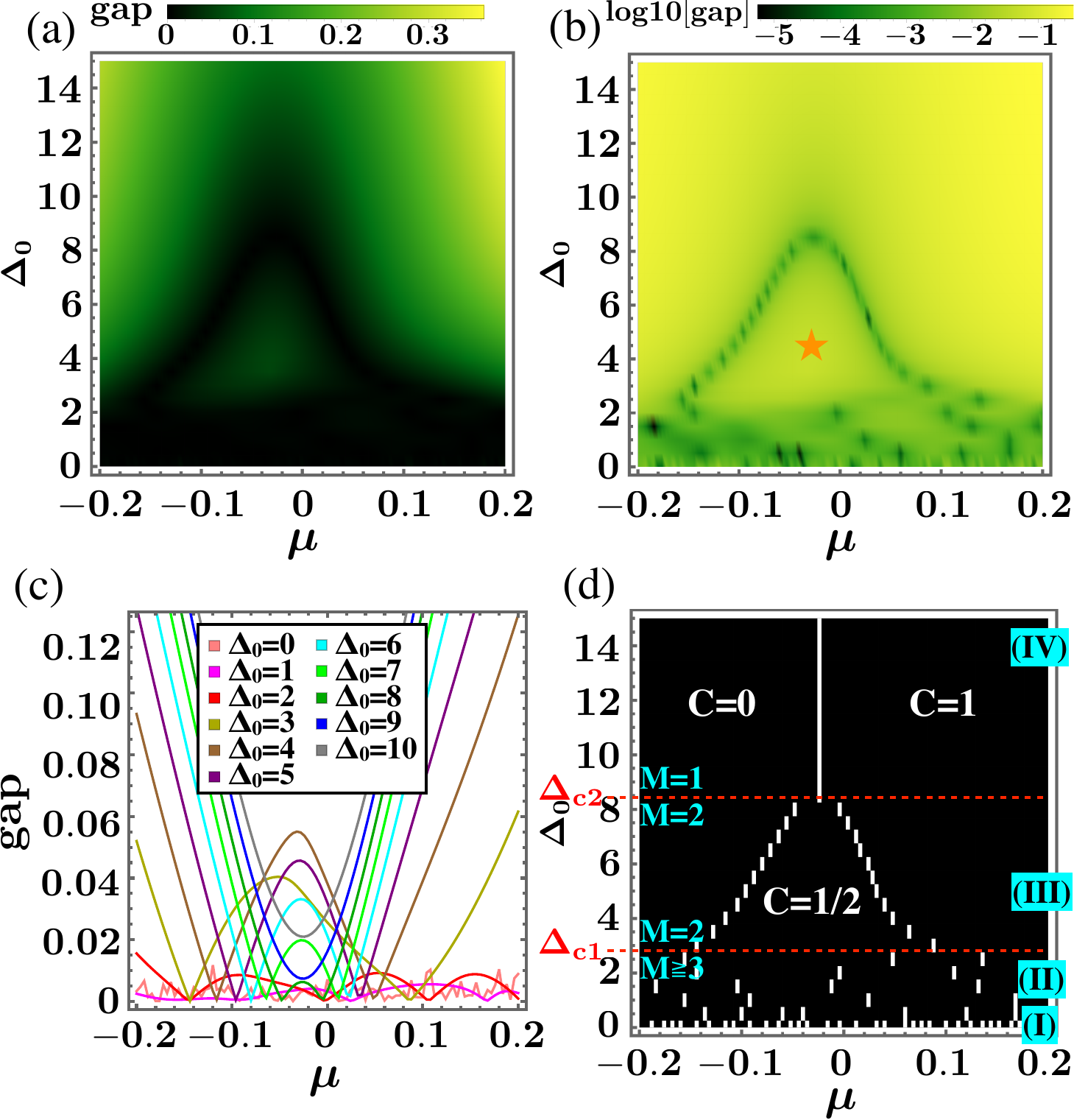}
 \caption{
 (a)(b) Density plots of the energy gap and its logarithm as functions of 
 the chemical potential $\mu$ and the pairing amplitude $\Delta_0$. The 
 star in (b) indicates a topological superconducting phase.
 (c) Energy gap as a function of $\mu$. 
 (d) White points indicate the local-minimum gap. When identifying these 
 points, we sweep $\mu$ in increments of $\delta\mu=0.004$ at each $\Delta_0$. 
 The red dashed lines represent $\Delta_\text{c1}$ and $\Delta_\text{c2}$, 
 representing the boundary where the number of local-minimum gap points, $M$, 
 changes. There are four regions as labeled by
 \eqref{item:1}-\eqref{item:4}.
 We set $2Q=198$ and $\Gamma=1$ in all of the figures.
 }
 \label{fig:gap}
\end{figure}

\subsection{Energy gap}
Figure~\ref{fig:gap}(a) plots the energy gap ---the energy separation around 
$E=0$ in the BdG spectrum--- as a function of $\mu$ and $\Delta_0$ with 
$2Q=198$, which is a ``typical'' value of $2Q$ as mentioned below. 
To visualize gap-closing points, we present its logarithm in 
Fig.~\ref{fig:gap}(b). The subsequent analysis suggests that the centered 
phase with a star in the figure 
is a topological superconducting phase. Figure~\ref{fig:gap}(c) explicitly 
shows the values of the gap at specific $\Delta_0$'s, indicating that each 
local minimum corresponds to the gap-closing as long as 
$0<\Delta_0\lesssim8.5$.

To clarify emergent phases, we generate Fig.~\ref{fig:gap}(d) that 
displays local-minimum gap points. Here, we count the number of those points 
at each $\Delta_0$, denoted $M$, and then define $\Delta_\text{c1}$ 
($\Delta_\text{c2}$) as the transition points where $M$ changes to 2 (1) as 
$\Delta_0$ increases. In Fig.~\ref{fig:gap}(d), we have $\Delta_\text{c1}\sim3$
and $\Delta_\text{c2}\sim8.5$. Let us now discuss each of the following four 
regions:
(The ground state at $|\mu|\gg\Delta_0,\Gamma$ is completely occupied or 
unoccupied states. Associating the Chern number $C$, we call each of them 
``$C=1$'' and ``$C=0$'' states, respectively.

\begin{enumerate}[(I):]
 \setlength{\parskip}{0cm}
 \setlength{\itemsep}{0cm}
 \item $\Delta_0=0$: {\it QH plateau transition.}
       
       A plateau transition connecting $C=0$ and $1$ phases occurs as $\mu$ 
       is changed. 
       Localized states induce multiple gap-closing points in 
       Fig.~\ref{fig:gap}(d).
       \label{item:1}
 
 \item $0<\Delta_0<\Delta_\text{c1}$: {\it Open question.}
       
       We have $M\geq3$.
       The nature of the gap-closing points is still an open question.
       There are two possible scenarios: a plateau transition connecting 
       (1) $C=0,1/2$ and $1$ phases,
       or (2) $C=0$ and $1$ phases. The distinction lies in whether the 
       intermediate state at $\mu\sim0$ is a topological SC (see the next 
       paragraph for more details) or a gapless state~\cite{Mishmash19} arising
       from a network model with states carrying different Chern numbers.
       Determining  the possibilities requires more careful 
       calculations,
       which we leave for future study.
       
 \item $\Delta_\text{c1}<\Delta_0<\Delta_\text{c2}$: 
       {\it Emergence of $C=1/2$ phase.}
       
       We have $M=2$. 
       Given that the intermediate phase is situated between $C=0$ and $C=1$ 
       phases, we expect it to be the $C=1/2$ phase, namely
       a topological superconducting phase with the unit BdG Chern 
       number. This expectation is further supported by
       the entanglement spectrum calculated in the next section. 
       
 \item $\Delta_\text{c2}<\Delta_0$: 
       {\it Disappearance of $C=1/2$ phase.} 
       
       We have $M=1$. The value of the local minimum increases as $\Delta_0$ is
       increased, see Fig.~\ref{fig:gap}(c). 
       According to Ref.~\onlinecite{Mishmash19}, the pairing amplitude 
       induced from a mixed state SC
       contains 
       point nodes akin to the $p+ip$-wave pairing potential, which
       gives rise only two phases $C=0$ 
       and $C=1$. 
       This suggests that the observed finite minimum gap at $\Delta_0=9,10$
       is a finite-size effect, and therefore the line in Fig.~\ref{fig:gap}(d)
       can be interpreted 
       as the phase boundary dividing $C=0$ and $1$.
       \label{item:4}
       
\end{enumerate}

\subsection{Entanglement spectrum}
Now we topologically characterize the $C=1/2$ phase. Although a typical way is 
to calculate the Chern number, but is fundamentally challenging in our model 
due to
the absence of translation invariance in the spherical geometry. 
Instead, 
we calculate the single-particle real-space entanglement spectrum (EtS), which 
is an energy-like 
spectrum with virtual open boundary conditions~\cite{Ryu06,Li08,Prodan10}.
By identifying gapless modes in EtS, we diagnose topological nature in each
phase.

The EtS is determined from the properties of the correlation functions. We first 
review this using a quadratic BdG Hamiltonian as 
$(1/2)\sum_{ij}\bm{a}^\dagger_iH_{ij}\bm{a}_j$ with
$\bm{a}_i^\dagger=(a_i^\dagger,a_i)$ on a lattice, where spin or orbital 
indices can be 
added along with the spatial coordinate $i$. By dividing the system 
into to two parts, $A$ and $B$, the entanglement Hamiltonian $\mathcal{H}_A$ 
is defined by 
\begin{align}
 \rho_A\equiv\tr_B\rho=\frac{1}{Z}e^{-\mathcal{H}_A},
\end{align}
where $\rho=\ket{G}\bra{G}$ with $\ket{G}$ the ground state, $\tr_B$ refers 
to the trace over the region $B$, and $Z$ is a normalization constant. The
entanglement Hamiltonian also has a quadratic form as
$\mathcal{H}_A
=(1/2)\sum_{ij\in A}\bm{a}^\dagger_i(H_A)_{ij}\bm{a}_j$~\cite{Peschel03}.
The matrix $H_A$ can be expressed using the correlation function matrix
$(C_A)_{i,j\in A}=\tr\left[\rho_A\bm{a}_i\bm{a}_j^\dagger\right]$
as~\cite{Cheong04,Peschel03,Oliveira14}
\begin{align}
 C_A=\frac{1}{1+e^{-H_A}}.
\end{align}
Since the spectra of $H_A$ and $C_A$ have the qualitatively
same structure, we
focus on $C_A$ and refer to its eigenvalues $\zeta_k$'s as the EtS below. The 
particle-hole symmetry $PH_AP^{-1}=-H_A$ leads to $PC_AP^{-1}=1-C_A$, where $P$
is an anti-unitary matrix. Therefore, the EtS is symmetric with respect to 
$1/2$ because of $0\leq\zeta_k\leq1$.



Assuming the validity of applying the above discussion to a continuum system, 
we calculate the correlation function only on the northern hemisphere 
(NH) as
\begin{align}
 \begin{split}
  C_\text{NH}(\vec{\Omega},\vec{\Omega}')
  &\equiv\int_\text{NH}d\vec{\Omega}\,
  \bra{G}
  \bm{c}(\vec{\Omega})\bm{c}^\dagger(\vec{\Omega})
  \ket{G}\\
  &=\delta(\vec{\Omega},\vec{\Omega}')
  +Y(\vec{\Omega})AY^\dagger(\vec{\Omega}'),
 \end{split} 
 \label{eq:CNH}
\end{align}
where $\bm{c}^\dagger
=(c^\dagger_\uparrow,c^\dagger_\downarrow,c_\uparrow,c_\downarrow)$.
The matrices in the second line in Eq.~\eqref{eq:CNH} are defined as follows:
\begin{align}
 \delta(\vec{\Omega},\vec{\Omega}')
 &\equiv
 \left(
 \begin{array}{cccc}
  \delta^2(\vec{\Omega}-\vec{\Omega}') \\
  & \delta^2(\vec{\Omega}-\vec{\Omega}') \\
  & & 0 \\
  & & & 0
 \end{array}
 \right),\\
 Y
 &=\left(
 \begin{array}{cccc}
  \bm{Y}_{Q_+Q_+} \\
  & \bm{Y}_{Q_-Q_+} \\
  & & \bm{Y}^*_{Q_+Q_+} \\
  & & & \bm{Y}^*_{Q_-Q_+} \\
 \end{array}
 \right)\\
 A&=\left(
 \begin{array}{cc}
  -M_uM_u^\dagger & -M_uM_v^\dagger \\
  M_u^*M_v^T & M_u^*M_u^T \\
 \end{array}
 \right),
\end{align}
where 
\begin{align}
 &\bm{Y}_{Ql}(\vec{\Omega})
 =(Y_{Ql-l}(\vec{\Omega}),\ldots,Y_{Qll}(\vec{\Omega})),\\
 &M_u=\vec{v}_{1-}\otimes u,\\
 &M_v=\vec{v}_{1-}^*\otimes v.
\end{align} 
Here, $\otimes$ refers to the tensor product. Note that $(u,v)$ has been 
defined in Eq.~\eqref{eq:uv} above.

The EtS $\zeta_k$ is given by the eigenvalue problem,
\begin{align}
 \int_\text{NH}d\vec{\Omega}'
 C(\vec{\Omega},\vec{\Omega}')\vec{\psi}_k(\vec{\Omega}')
 =\zeta_k\vec{\psi}_k(\vec{\Omega}).
 \label{eq:ev1}
\end{align}
By expanding $\vec{\psi}_k(\vec{\Omega})=Y(\vec{\Omega})\vec{\alpha}_k$
and 
operating $\int d\vec{\Omega}\,Y^\dagger(\vec{\Omega})$ from the left, 
Eq.~\eqref{eq:ev1}
reduces to
\begin{align}
 &\left(\delta'-AB\right)\vec{\alpha}_k=\zeta_k\vec{\alpha}_k,
 \label{eq:ev2}
\end{align}
where 
\begin{align}
 &B=\int_\text{NH}d\vec{\Omega}\vec{Y}^\dagger\vec{Y},\\
 &\delta'=\text{diag}\{\bm{1}_{4Q_++2},\bm0_{4Q_++2}\},
\end{align}
 with $\bm{1}_n$
($\bm{0}_n$) the $n$-dimensional identity (zero) matrices. Although 
$\delta'-AB$ is not hermitian, Eq.~\eqref{eq:ev2} can be transformed to an
Hermitian problem as
\begin{align}
 \left(\delta'-B^{1/2}AB^{1/2}\right)\vec{\alpha}_k'=\zeta_k\vec{\alpha}_k',
\end{align}
where $\vec{\alpha}_k'=B^{1/2}\vec{\alpha}_k$. We note that $A$ is Hermitian, 
and $B=\text{diag}\{B_\uparrow,B_\downarrow,B_\uparrow,B_\downarrow\}$
with $\left(B_\sigma\right)_{mm'}=\delta_{mm'}\int_\text{NH}d\vec{\Omega}
\left|Y_{Q_{\pm}Q_+m}\right|^2$
is positive-definite (This integration reduces to the Beta function). 
While one obtains $\zeta_k=0$ or 1 at $\mu\rightarrow-\infty$,
the ground state at $\mu\rightarrow\infty$, i.e. the IQH state, produces
``entanglement gapless modes'' that cross $\zeta_k=0$ and 
1~\cite{Rodriguez09}. Below, we explore the 
EtS in the intermediate range of $\mu$.

\begin{figure}
 \includegraphics[width=\columnwidth]{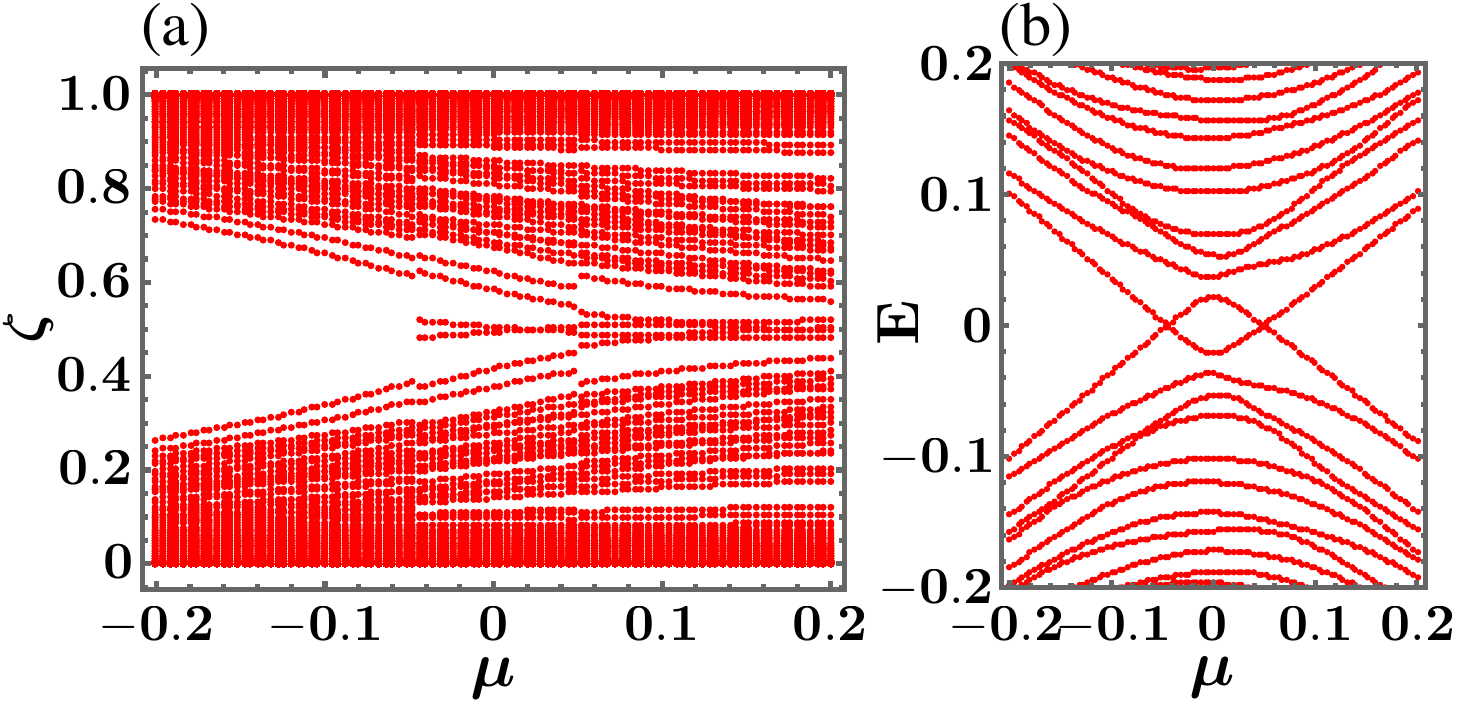}
 \caption{
 (a) EtS and (b) BdG spectrum at $\Delta_0=4$ with $2Q=222$. We set
 $\Gamma=1$.
 }
 \label{fig:EtS}
\end{figure}
Figure~\ref{fig:EtS}(a) illustrates the EtS at $\Delta_0=4$ as a function of 
the chemical potential $\mu$. In comparison, Fig.~\ref{fig:EtS}(b) shows the 
BdG spectrum, revealing two gap-closing points at $\mu\sim\pm0.05$. The mid-gap
states suddenly appear in the EtS at $\mu\sim-0.05$, and their density 
increases discretely at $\mu\sim0.05$. This 
observation is consistent with the existence of the three phases, namely, 
the trivial ($C=0$), the topological superconducting ($C=1/2$), and the IQH 
($C=1$) phases. In the figures, we set $2Q=222$ because this system has the 
relatively large energy gap of the ground state, resulting in less finite-size 
effect in the EtS.

\subsection{Ensemble average}
So far we have discussed findings based on a single impurity distribution. In 
the remainder of the work, we will present results using multiple random 
distributions.

\begin{figure}[t]
 \includegraphics[width=\columnwidth]{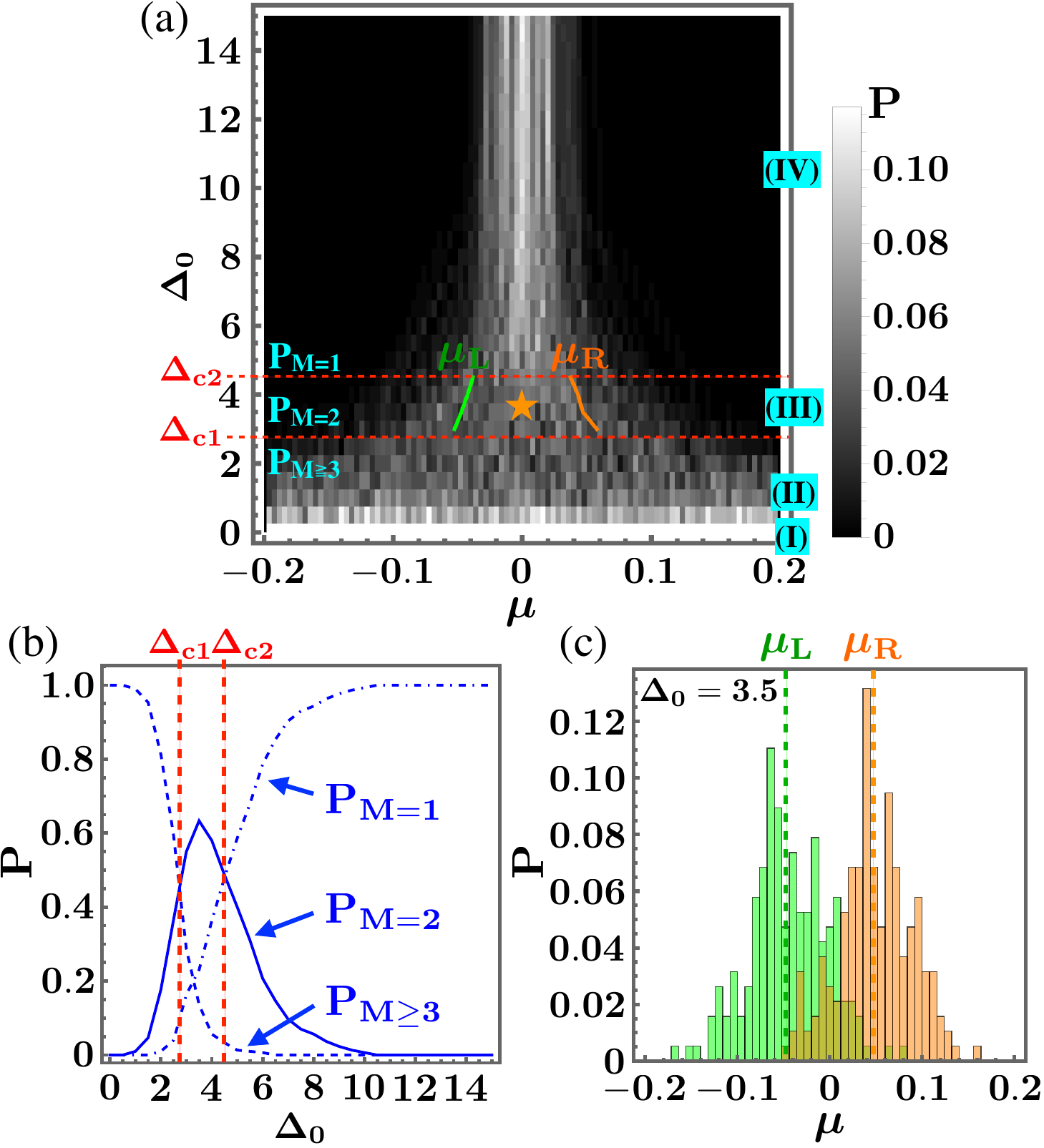}
 \caption{
 (a) Distribution of local-minimum gap points. The color indicates the 
 normalized count. The star denotes the topological
 superconducting regime, enclosed by two red-dashed lines (representing 
 $\Delta_\text{c1}$ and $\Delta_\text{c2}$) and green and orange line 
 (representing $\mu_L$ and $\mu_R$). 
 (b) Probability of having $M$ local-minimum gap points. The values of 
 $\Delta_\text{c1}$ and $\Delta_\text{c2}$ are deduced from intersections of 
 each line.
 (c) Distribution of local-minimum gap points with $M=2$ at $\Delta_0=3.5$.
 The green and orange colors represent data for the left and right 
 local-minimum gap, respectively. The values of $\mu_L$ and $\mu_R$ are defined
 as means for each dataset. We set $2Q=198$ and $\Gamma=1$ in all of the 
 figures.
 }
 \label{fig:histgram}
\end{figure}

Figure~\ref{fig:histgram}(a) is an analog of Fig.~\ref{fig:gap}(d) but 
incorporates 300 random distributions of impurities. This is a 2D histogram 
where the gray color represents the normalized count
of local-mininum-gap samples at each point. 
The superconducting phase is expected to emerge in 
$\Delta_\text{1c}<\Delta_0<\Delta_\text{2c}$ and 
$\mu_\text{L}<\mu<\mu_\text{R}$ as indicated by a star. Each phase boundary
is identified as follows: Figure~\ref{fig:histgram}(b) plots the probability of
having $M$ local-minimum gap points, denoted $P_M$, as a function of 
$\Delta_0$. We define $\Delta_\text{c1}$ ($\Delta_\text{c2}$) by 
intersections of $P_{M\geq3}$ and $P_{M=2}$ ($P_{M=2}$ and $P_{M=1}$). We
determine $\mu_R$ and $\mu_L$ using histograms as in 
Fig.~\ref{fig:histgram}(c), where the green (orange) bars represent the 
distribution of left (right) local-minimum gap points. We define $\mu_L$ and 
$\mu_R$ as means for each dataset. 

\begin{figure}[t!]
 \includegraphics[width=\columnwidth]{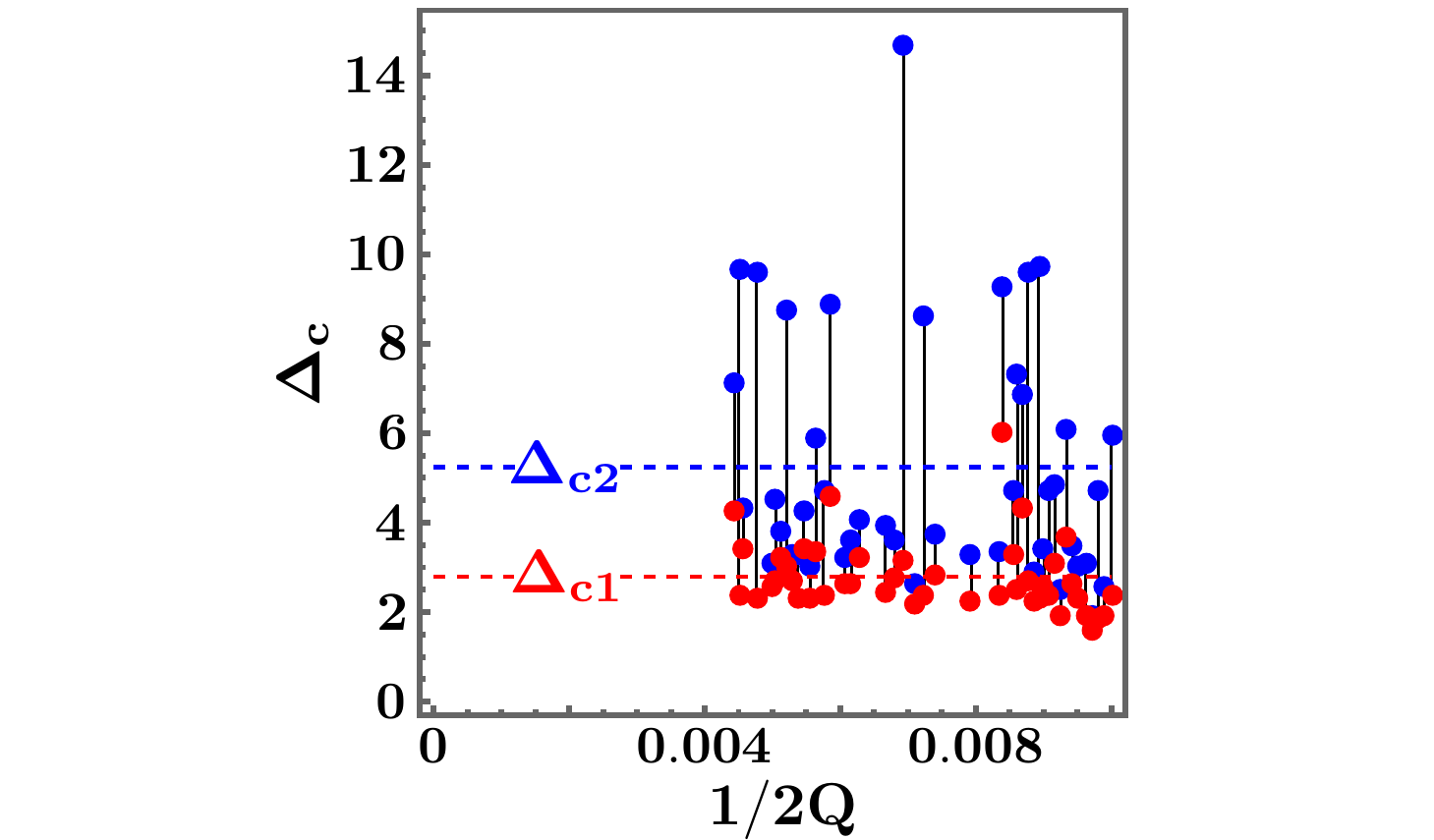}
 \caption{
 Size-scaling analysis for $\Delta_\text{c1}$ (red dots) and 
$\Delta_\text{c2}$ (blue dots). The black
 lines connect the red and the blue dots at each $2Q$. The dashed lines 
 represent the average of each dataset.
 }
 \label{fig:average}
\end{figure}
To explore the thermodynamic limit, we calculate $\Delta_\text{c1}$ and 
$\Delta_\text{c2}$ for different $2Q$'s in the above way, 
and present the results in Fig.~\ref{fig:average}. The dashed lines indicate 
the mean values of each dataset. The difference 
$\Delta_\text{c2}-\Delta_\text{c1}$ appears to remain finite as systems 
increase in size, suggesting the presence of the topological superconducting 
phase in the thermodynamic limit. We have chosen $2Q=198$ 
in the above analysis as a representative example because it is the largest 
$2Q$ within the five closest systems to the two dashed lines.

\section{Concluding remarks}
\label{sec:conc}
In this paper, we formulate a scheme to combine the Rashba-coupled QH
system with a type-II $s$-wave SC on the spherical geometry. Through this 
setup, we numerically demonstrate the emergence of a disorder-induced 
phase between a trivial and the IQH phases.
We expect this to be the superconducting phase associated with
the half-integer Chern number. This expectation is further supported by 
revealing the entanglement gapless modes in the EtS.

Our spherical model will be a useful platform for future study on topological
order. Although we have focused on the (Rashba-coupled) IQH system, our 
formulation is also 
valid for the FQH system. The interplay between fractionalization and 
superconductivity can give rise to intriguing phenomena such as $\mathbb{Z}_n$ 
parafermions or Fibonacci anyons as mentioned in the introduction. 
The topological degeneracy of such states, stemming from their non-Abelian 
nature, is
generally not exact in a finite system, which can lead to subtle problems in, 
for example, assessing degeneracy or energy excitations in a translationally 
invariant system. Our spherical model 
inherently avoids this issue, facilitating the straightforward identification 
of non-Abelian topological order.

\begin{acknowledgments}
 We acknowledge the computational resources offered by Research Institute for
 Information Technology, Kyushu University. The work is supported in part by 
 JSPS KAKENHI Grant no. JP23K19036, 
 JP20H01830, 
 and JST CREST Grant
 no. JPMJCR18T2. 
\end{acknowledgments}

\bibliography{biblio_fqhe}

\begin{thebibliography}{111}
\expandafter\ifx\csname natexlab\endcsname\relax\def\natexlab#1{#1}\fi
\expandafter\ifx\csname bibnamefont\endcsname\relax
  \def\bibnamefont#1{#1}\fi
\expandafter\ifx\csname bibfnamefont\endcsname\relax
  \def\bibfnamefont#1{#1}\fi
\expandafter\ifx\csname citenamefont\endcsname\relax
  \def\citenamefont#1{#1}\fi
\expandafter\ifx\csname url\endcsname\relax
  \def\url#1{\texttt{#1}}\fi
\expandafter\ifx\csname urlprefix\endcsname\relax\def\urlprefix{URL }\fi
\providecommand{\bibinfo}[2]{#2}
\providecommand{\eprint}[2][]{\url{#2}}

\bibitem[{\citenamefont{Wu}(1984)}]{Wu84b}
\bibinfo{author}{\bibfnamefont{Y.-S.} \bibnamefont{Wu}},
  \bibinfo{journal}{Phys. Rev. Lett.} \textbf{\bibinfo{volume}{52}},
  \bibinfo{pages}{2103} (\bibinfo{year}{1984}),
  \urlprefix\url{https://link.aps.org/doi/10.1103/PhysRevLett.52.2103}.

\bibitem[{\citenamefont{Wilczek}(1982)}]{Wilczek82}
\bibinfo{author}{\bibfnamefont{F.}~\bibnamefont{Wilczek}},
  \bibinfo{journal}{Phys. Rev. Lett.} \textbf{\bibinfo{volume}{49}},
  \bibinfo{pages}{957} (\bibinfo{year}{1982}),
  \urlprefix\url{http://link.aps.org/doi/10.1103/PhysRevLett.49.957}.

\bibitem[{\citenamefont{Wen}(1995)}]{Wen95}
\bibinfo{author}{\bibfnamefont{X.-G.} \bibnamefont{Wen}},
  \bibinfo{journal}{Advances in Physics} \textbf{\bibinfo{volume}{44}},
  \bibinfo{pages}{405} (\bibinfo{year}{1995}),
  \eprint{http://www.tandfonline.com/doi/pdf/10.1080/00018739500101566},
  \urlprefix\url{http://www.tandfonline.com/doi/abs/10.1080/00018739500101566}.

\bibitem[{\citenamefont{Tsui et~al.}(1982)\citenamefont{Tsui, Stormer, and
  Gossard}}]{Tsui82}
\bibinfo{author}{\bibfnamefont{D.~C.} \bibnamefont{Tsui}},
  \bibinfo{author}{\bibfnamefont{H.~L.} \bibnamefont{Stormer}},
  \bibnamefont{and} \bibinfo{author}{\bibfnamefont{A.~C.}
  \bibnamefont{Gossard}}, \bibinfo{journal}{Phys. Rev. Lett.}
  \textbf{\bibinfo{volume}{48}}, \bibinfo{pages}{1559} (\bibinfo{year}{1982}),
  \urlprefix\url{http://link.aps.org/doi/10.1103/PhysRevLett.48.1559}.

\bibitem[{\citenamefont{Laughlin}(1983)}]{Laughlin83}
\bibinfo{author}{\bibfnamefont{R.~B.} \bibnamefont{Laughlin}},
  \bibinfo{journal}{Phys. Rev. Lett.} \textbf{\bibinfo{volume}{50}},
  \bibinfo{pages}{1395} (\bibinfo{year}{1983}),
  \urlprefix\url{http://link.aps.org/doi/10.1103/PhysRevLett.50.1395}.

\bibitem[{\citenamefont{Jain}(2007)}]{Jain07}
\bibinfo{author}{\bibfnamefont{J.~K.} \bibnamefont{Jain}},
  \emph{\bibinfo{title}{Composite Fermions}} (\bibinfo{publisher}{Cambridge
  University Press, New York, US}, \bibinfo{year}{2007}).

\bibitem[{\citenamefont{Kalmeyer and Laughlin}(1987)}]{Kalmeyer87}
\bibinfo{author}{\bibfnamefont{V.}~\bibnamefont{Kalmeyer}} \bibnamefont{and}
  \bibinfo{author}{\bibfnamefont{R.~B.} \bibnamefont{Laughlin}},
  \bibinfo{journal}{Phys. Rev. Lett.} \textbf{\bibinfo{volume}{59}},
  \bibinfo{pages}{2095} (\bibinfo{year}{1987}),
  \urlprefix\url{https://link.aps.org/doi/10.1103/PhysRevLett.59.2095}.

\bibitem[{\citenamefont{Levin and Wen}(2005)}]{Levin05a}
\bibinfo{author}{\bibfnamefont{M.~A.} \bibnamefont{Levin}} \bibnamefont{and}
  \bibinfo{author}{\bibfnamefont{X.-G.} \bibnamefont{Wen}},
  \bibinfo{journal}{Phys. Rev. B} \textbf{\bibinfo{volume}{71}},
  \bibinfo{pages}{045110} (\bibinfo{year}{2005}),
  \urlprefix\url{http://link.aps.org/doi/10.1103/PhysRevB.71.045110}.

\bibitem[{\citenamefont{Kitaev}(2006)}]{Kitaev06}
\bibinfo{author}{\bibfnamefont{A.}~\bibnamefont{Kitaev}},
  \bibinfo{journal}{Annals of Physics} \textbf{\bibinfo{volume}{321}},
  \bibinfo{pages}{2 } (\bibinfo{year}{2006}), ISSN \bibinfo{issn}{0003-4916},
  \bibinfo{note}{january Special Issue},
  \urlprefix\url{http://www.sciencedirect.com/science/article/pii/S0003491605002381}.

\bibitem[{\citenamefont{Jackeli and Khaliullin}(2009)}]{Jackeli09}
\bibinfo{author}{\bibfnamefont{G.}~\bibnamefont{Jackeli}} \bibnamefont{and}
  \bibinfo{author}{\bibfnamefont{G.}~\bibnamefont{Khaliullin}},
  \bibinfo{journal}{Phys. Rev. Lett.} \textbf{\bibinfo{volume}{102}},
  \bibinfo{pages}{017205} (\bibinfo{year}{2009}),
  \urlprefix\url{https://link.aps.org/doi/10.1103/PhysRevLett.102.017205}.

\bibitem[{\citenamefont{Read and Green}(2000)}]{Read00}
\bibinfo{author}{\bibfnamefont{N.}~\bibnamefont{Read}} \bibnamefont{and}
  \bibinfo{author}{\bibfnamefont{D.}~\bibnamefont{Green}},
  \bibinfo{journal}{Phys. Rev. B} \textbf{\bibinfo{volume}{61}},
  \bibinfo{pages}{10267} (\bibinfo{year}{2000}),
  \urlprefix\url{http://link.aps.org/doi/10.1103/PhysRevB.61.10267}.

\bibitem[{\citenamefont{Ivanov}(2001)}]{Ivanov01}
\bibinfo{author}{\bibfnamefont{D.~A.} \bibnamefont{Ivanov}},
  \bibinfo{journal}{Phys. Rev. Lett.} \textbf{\bibinfo{volume}{86}},
  \bibinfo{pages}{268} (\bibinfo{year}{2001}),
  \urlprefix\url{http://link.aps.org/doi/10.1103/PhysRevLett.86.268}.

\bibitem[{\citenamefont{Kasahara et~al.}(2018)\citenamefont{Kasahara, Ohnishi,
  Mizukami, Tanaka, Ma, Sugii, Kurita, Tanaka, Nasu, Motome
  et~al.}}]{Kasahara18}
\bibinfo{author}{\bibfnamefont{Y.}~\bibnamefont{Kasahara}},
  \bibinfo{author}{\bibfnamefont{T.}~\bibnamefont{Ohnishi}},
  \bibinfo{author}{\bibfnamefont{Y.}~\bibnamefont{Mizukami}},
  \bibinfo{author}{\bibfnamefont{O.}~\bibnamefont{Tanaka}},
  \bibinfo{author}{\bibfnamefont{S.}~\bibnamefont{Ma}},
  \bibinfo{author}{\bibfnamefont{K.}~\bibnamefont{Sugii}},
  \bibinfo{author}{\bibfnamefont{N.}~\bibnamefont{Kurita}},
  \bibinfo{author}{\bibfnamefont{H.}~\bibnamefont{Tanaka}},
  \bibinfo{author}{\bibfnamefont{J.}~\bibnamefont{Nasu}},
  \bibinfo{author}{\bibfnamefont{Y.}~\bibnamefont{Motome}},
  \bibnamefont{et~al.}, \bibinfo{journal}{Nature}
  \textbf{\bibinfo{volume}{559}}, \bibinfo{pages}{227} (\bibinfo{year}{2018}),
  \urlprefix\url{https://doi.org/10.1038/s41586-018-0274-0}.

\bibitem[{\citenamefont{Yokoi et~al.}(2021)\citenamefont{Yokoi, Ma, Kasahara,
  Kasahara, Shibauchi, Kurita, Tanaka, Nasu, Motome, Hickey et~al.}}]{Yokoi21}
\bibinfo{author}{\bibfnamefont{T.}~\bibnamefont{Yokoi}},
  \bibinfo{author}{\bibfnamefont{S.}~\bibnamefont{Ma}},
  \bibinfo{author}{\bibfnamefont{Y.}~\bibnamefont{Kasahara}},
  \bibinfo{author}{\bibfnamefont{S.}~\bibnamefont{Kasahara}},
  \bibinfo{author}{\bibfnamefont{T.}~\bibnamefont{Shibauchi}},
  \bibinfo{author}{\bibfnamefont{N.}~\bibnamefont{Kurita}},
  \bibinfo{author}{\bibfnamefont{H.}~\bibnamefont{Tanaka}},
  \bibinfo{author}{\bibfnamefont{J.}~\bibnamefont{Nasu}},
  \bibinfo{author}{\bibfnamefont{Y.}~\bibnamefont{Motome}},
  \bibinfo{author}{\bibfnamefont{C.}~\bibnamefont{Hickey}},
  \bibnamefont{et~al.}, \bibinfo{journal}{Science}
  \textbf{\bibinfo{volume}{373}}, \bibinfo{pages}{568} (\bibinfo{year}{2021}),
  \eprint{https://www.science.org/doi/pdf/10.1126/science.aay5551},
  \urlprefix\url{https://www.science.org/doi/abs/10.1126/science.aay5551}.

\bibitem[{\citenamefont{Bruin et~al.}(2022)\citenamefont{Bruin, Claus,
  Matsumoto, Kurita, Tanaka, and Takagi}}]{Bruin22}
\bibinfo{author}{\bibfnamefont{J.~A.~N.} \bibnamefont{Bruin}},
  \bibinfo{author}{\bibfnamefont{R.~R.} \bibnamefont{Claus}},
  \bibinfo{author}{\bibfnamefont{Y.}~\bibnamefont{Matsumoto}},
  \bibinfo{author}{\bibfnamefont{N.}~\bibnamefont{Kurita}},
  \bibinfo{author}{\bibfnamefont{H.}~\bibnamefont{Tanaka}}, \bibnamefont{and}
  \bibinfo{author}{\bibfnamefont{H.}~\bibnamefont{Takagi}},
  \bibinfo{journal}{Nature Physics} \textbf{\bibinfo{volume}{18}},
  \bibinfo{pages}{401} (\bibinfo{year}{2022}),
  \urlprefix\url{https://doi.org/10.1038/s41567-021-01501-y}.

\bibitem[{\citenamefont{Nakamura et~al.}(2020)\citenamefont{Nakamura, Liang,
  Gardner, and Manfra}}]{Nakamura20}
\bibinfo{author}{\bibfnamefont{J.}~\bibnamefont{Nakamura}},
  \bibinfo{author}{\bibfnamefont{S.}~\bibnamefont{Liang}},
  \bibinfo{author}{\bibfnamefont{G.~C.} \bibnamefont{Gardner}},
  \bibnamefont{and} \bibinfo{author}{\bibfnamefont{M.~J.}
  \bibnamefont{Manfra}}, \bibinfo{journal}{Nature Physics}
  \textbf{\bibinfo{volume}{16}}, \bibinfo{pages}{931} (\bibinfo{year}{2020}),
  \urlprefix\url{https://doi.org/10.1038/s41567-020-1019-1}.

\bibitem[{\citenamefont{Bartolomei et~al.}(2020)\citenamefont{Bartolomei,
  Kumar, Bisognin, Marguerite, Berroir, Bocquillon, Pla^^c3^^a7ais, Cavanna,
  Dong, Gennser et~al.}}]{Bartolomei20}
\bibinfo{author}{\bibfnamefont{H.}~\bibnamefont{Bartolomei}},
  \bibinfo{author}{\bibfnamefont{M.}~\bibnamefont{Kumar}},
  \bibinfo{author}{\bibfnamefont{R.}~\bibnamefont{Bisognin}},
  \bibinfo{author}{\bibfnamefont{A.}~\bibnamefont{Marguerite}},
  \bibinfo{author}{\bibfnamefont{J.-M.} \bibnamefont{Berroir}},
  \bibinfo{author}{\bibfnamefont{E.}~\bibnamefont{Bocquillon}},
  \bibinfo{author}{\bibfnamefont{B.}~\bibnamefont{Pla^^c3^^a7ais}},
  \bibinfo{author}{\bibfnamefont{A.}~\bibnamefont{Cavanna}},
  \bibinfo{author}{\bibfnamefont{Q.}~\bibnamefont{Dong}},
  \bibinfo{author}{\bibfnamefont{U.}~\bibnamefont{Gennser}},
  \bibnamefont{et~al.}, \bibinfo{journal}{Science}
  \textbf{\bibinfo{volume}{368}}, \bibinfo{pages}{173} (\bibinfo{year}{2020}),
  \eprint{https://www.science.org/doi/pdf/10.1126/science.aaz5601},
  \urlprefix\url{https://www.science.org/doi/abs/10.1126/science.aaz5601}.

\bibitem[{\citenamefont{Nakamura et~al.}(2023)\citenamefont{Nakamura, Liang,
  Gardner, and Manfra}}]{Nakamura23}
\bibinfo{author}{\bibfnamefont{J.}~\bibnamefont{Nakamura}},
  \bibinfo{author}{\bibfnamefont{S.}~\bibnamefont{Liang}},
  \bibinfo{author}{\bibfnamefont{G.~C.} \bibnamefont{Gardner}},
  \bibnamefont{and} \bibinfo{author}{\bibfnamefont{M.~J.}
  \bibnamefont{Manfra}}, \emph{\bibinfo{title}{Fabry-perot interferometry at
  the $\nu$ = 2/5 fractional quantum hall state}} (\bibinfo{year}{2023}),
  \eprint{2304.12415}.

\bibitem[{\citenamefont{Kitaev}(2003)}]{Kitaev03}
\bibinfo{author}{\bibfnamefont{A.}~\bibnamefont{Kitaev}},
  \bibinfo{journal}{Annals of Physics} \textbf{\bibinfo{volume}{303}},
  \bibinfo{pages}{2 } (\bibinfo{year}{2003}), ISSN \bibinfo{issn}{0003-4916},
  \urlprefix\url{http://www.sciencedirect.com/science/article/pii/S0003491602000180}.

\bibitem[{\citenamefont{Nayak et~al.}(2008)\citenamefont{Nayak, Simon, Stern,
  Freedman, and Das~Sarma}}]{Nayak08}
\bibinfo{author}{\bibfnamefont{C.}~\bibnamefont{Nayak}},
  \bibinfo{author}{\bibfnamefont{S.~H.} \bibnamefont{Simon}},
  \bibinfo{author}{\bibfnamefont{A.}~\bibnamefont{Stern}},
  \bibinfo{author}{\bibfnamefont{M.}~\bibnamefont{Freedman}}, \bibnamefont{and}
  \bibinfo{author}{\bibfnamefont{S.}~\bibnamefont{Das~Sarma}},
  \bibinfo{journal}{Rev. Mod. Phys.} \textbf{\bibinfo{volume}{80}},
  \bibinfo{pages}{1083} (\bibinfo{year}{2008}),
  \urlprefix\url{http://link.aps.org/doi/10.1103/RevModPhys.80.1083}.

\bibitem[{\citenamefont{Jain}(2020)}]{Jain20b}
\bibinfo{author}{\bibfnamefont{J.~K.} \bibnamefont{Jain}},
  \bibinfo{journal}{Current Science} \textbf{\bibinfo{volume}{119}},
  \bibinfo{pages}{430} (\bibinfo{year}{2020}).

\bibitem[{\citenamefont{Fu and Kane}(2008)}]{Fu08}
\bibinfo{author}{\bibfnamefont{L.}~\bibnamefont{Fu}} \bibnamefont{and}
  \bibinfo{author}{\bibfnamefont{C.~L.} \bibnamefont{Kane}},
  \bibinfo{journal}{Phys. Rev. Lett.} \textbf{\bibinfo{volume}{100}},
  \bibinfo{pages}{096407} (\bibinfo{year}{2008}),
  \urlprefix\url{https://link.aps.org/doi/10.1103/PhysRevLett.100.096407}.

\bibitem[{\citenamefont{Akhmerov et~al.}(2009)\citenamefont{Akhmerov, Nilsson,
  and Beenakker}}]{Akhmerov09}
\bibinfo{author}{\bibfnamefont{A.~R.} \bibnamefont{Akhmerov}},
  \bibinfo{author}{\bibfnamefont{J.}~\bibnamefont{Nilsson}}, \bibnamefont{and}
  \bibinfo{author}{\bibfnamefont{C.~W.~J.} \bibnamefont{Beenakker}},
  \bibinfo{journal}{Phys. Rev. Lett.} \textbf{\bibinfo{volume}{102}},
  \bibinfo{pages}{216404} (\bibinfo{year}{2009}),
  \urlprefix\url{https://link.aps.org/doi/10.1103/PhysRevLett.102.216404}.

\bibitem[{\citenamefont{Sau et~al.}(2010)\citenamefont{Sau, Lutchyn, Tewari,
  and Das~Sarma}}]{Sau10}
\bibinfo{author}{\bibfnamefont{J.~D.} \bibnamefont{Sau}},
  \bibinfo{author}{\bibfnamefont{R.~M.} \bibnamefont{Lutchyn}},
  \bibinfo{author}{\bibfnamefont{S.}~\bibnamefont{Tewari}}, \bibnamefont{and}
  \bibinfo{author}{\bibfnamefont{S.}~\bibnamefont{Das~Sarma}},
  \bibinfo{journal}{Phys. Rev. Lett.} \textbf{\bibinfo{volume}{104}},
  \bibinfo{pages}{040502} (\bibinfo{year}{2010}),
  \urlprefix\url{https://link.aps.org/doi/10.1103/PhysRevLett.104.040502}.

\bibitem[{\citenamefont{Lutchyn et~al.}(2010)\citenamefont{Lutchyn, Sau, and
  Das~Sarma}}]{Lutchyn10}
\bibinfo{author}{\bibfnamefont{R.~M.} \bibnamefont{Lutchyn}},
  \bibinfo{author}{\bibfnamefont{J.~D.} \bibnamefont{Sau}}, \bibnamefont{and}
  \bibinfo{author}{\bibfnamefont{S.}~\bibnamefont{Das~Sarma}},
  \bibinfo{journal}{Phys. Rev. Lett.} \textbf{\bibinfo{volume}{105}},
  \bibinfo{pages}{077001} (\bibinfo{year}{2010}),
  \urlprefix\url{https://link.aps.org/doi/10.1103/PhysRevLett.105.077001}.

\bibitem[{\citenamefont{Oreg et~al.}(2010)\citenamefont{Oreg, Refael, and
  Von~Oppen}}]{Oreg10}
\bibinfo{author}{\bibfnamefont{Y.}~\bibnamefont{Oreg}},
  \bibinfo{author}{\bibfnamefont{G.}~\bibnamefont{Refael}}, \bibnamefont{and}
  \bibinfo{author}{\bibfnamefont{F.}~\bibnamefont{Von~Oppen}},
  \bibinfo{journal}{Physical review letters} \textbf{\bibinfo{volume}{105}},
  \bibinfo{pages}{177002} (\bibinfo{year}{2010}).

\bibitem[{\citenamefont{Alicea}(2010)}]{Alicea10}
\bibinfo{author}{\bibfnamefont{J.}~\bibnamefont{Alicea}},
  \bibinfo{journal}{Phys. Rev. B} \textbf{\bibinfo{volume}{81}},
  \bibinfo{pages}{125318} (\bibinfo{year}{2010}),
  \urlprefix\url{https://link.aps.org/doi/10.1103/PhysRevB.81.125318}.

\bibitem[{\citenamefont{Qi et~al.}(2010)\citenamefont{Qi, Hughes, and
  Zhang}}]{Qi10}
\bibinfo{author}{\bibfnamefont{X.-L.} \bibnamefont{Qi}},
  \bibinfo{author}{\bibfnamefont{T.~L.} \bibnamefont{Hughes}},
  \bibnamefont{and} \bibinfo{author}{\bibfnamefont{S.-C.} \bibnamefont{Zhang}},
  \bibinfo{journal}{Phys. Rev. B} \textbf{\bibinfo{volume}{82}},
  \bibinfo{pages}{184516} (\bibinfo{year}{2010}),
  \urlprefix\url{https://link.aps.org/doi/10.1103/PhysRevB.82.184516}.

\bibitem[{\citenamefont{Zocher and Rosenow}(2016)}]{Bjorn16}
\bibinfo{author}{\bibfnamefont{B.}~\bibnamefont{Zocher}} \bibnamefont{and}
  \bibinfo{author}{\bibfnamefont{B.}~\bibnamefont{Rosenow}},
  \bibinfo{journal}{Phys. Rev. B} \textbf{\bibinfo{volume}{93}},
  \bibinfo{pages}{214504} (\bibinfo{year}{2016}),
  \urlprefix\url{https://link.aps.org/doi/10.1103/PhysRevB.93.214504}.

\bibitem[{\citenamefont{Mishmash et~al.}(2019)\citenamefont{Mishmash, Yazdani,
  and Zaletel}}]{Mishmash19}
\bibinfo{author}{\bibfnamefont{R.~V.} \bibnamefont{Mishmash}},
  \bibinfo{author}{\bibfnamefont{A.}~\bibnamefont{Yazdani}}, \bibnamefont{and}
  \bibinfo{author}{\bibfnamefont{M.~P.} \bibnamefont{Zaletel}},
  \bibinfo{journal}{Phys. Rev. B} \textbf{\bibinfo{volume}{99}},
  \bibinfo{pages}{115427} (\bibinfo{year}{2019}),
  \urlprefix\url{https://link.aps.org/doi/10.1103/PhysRevB.99.115427}.

\bibitem[{\citenamefont{Chaudhary and MacDonald}(2020)}]{Chaudhary20}
\bibinfo{author}{\bibfnamefont{G.}~\bibnamefont{Chaudhary}} \bibnamefont{and}
  \bibinfo{author}{\bibfnamefont{A.~H.} \bibnamefont{MacDonald}},
  \bibinfo{journal}{Phys. Rev. B} \textbf{\bibinfo{volume}{101}},
  \bibinfo{pages}{024516} (\bibinfo{year}{2020}),
  \urlprefix\url{https://link.aps.org/doi/10.1103/PhysRevB.101.024516}.

\bibitem[{\citenamefont{Cheng}(2012)}]{Cheng12}
\bibinfo{author}{\bibfnamefont{M.}~\bibnamefont{Cheng}},
  \bibinfo{journal}{Physical Review B} \textbf{\bibinfo{volume}{86}},
  \bibinfo{pages}{195126} (\bibinfo{year}{2012}).

\bibitem[{\citenamefont{Lindner et~al.}(2012)\citenamefont{Lindner, Berg,
  Refael, and Stern}}]{Lindner12}
\bibinfo{author}{\bibfnamefont{N.~H.} \bibnamefont{Lindner}},
  \bibinfo{author}{\bibfnamefont{E.}~\bibnamefont{Berg}},
  \bibinfo{author}{\bibfnamefont{G.}~\bibnamefont{Refael}}, \bibnamefont{and}
  \bibinfo{author}{\bibfnamefont{A.}~\bibnamefont{Stern}},
  \bibinfo{journal}{Physical Review X} \textbf{\bibinfo{volume}{2}},
  \bibinfo{pages}{041002} (\bibinfo{year}{2012}).

\bibitem[{\citenamefont{Burrello et~al.}(2013)\citenamefont{Burrello, van Heck,
  and Cobanera}}]{Burrello13}
\bibinfo{author}{\bibfnamefont{M.}~\bibnamefont{Burrello}},
  \bibinfo{author}{\bibfnamefont{B.}~\bibnamefont{van Heck}}, \bibnamefont{and}
  \bibinfo{author}{\bibfnamefont{E.}~\bibnamefont{Cobanera}},
  \bibinfo{journal}{Phys. Rev. B} \textbf{\bibinfo{volume}{87}},
  \bibinfo{pages}{195422} (\bibinfo{year}{2013}),
  \urlprefix\url{https://link.aps.org/doi/10.1103/PhysRevB.87.195422}.

\bibitem[{\citenamefont{Clarke et~al.}(2013)\citenamefont{Clarke, Alicea, and
  Shtengel}}]{Clarke13}
\bibinfo{author}{\bibfnamefont{D.~J.} \bibnamefont{Clarke}},
  \bibinfo{author}{\bibfnamefont{J.}~\bibnamefont{Alicea}}, \bibnamefont{and}
  \bibinfo{author}{\bibfnamefont{K.}~\bibnamefont{Shtengel}},
  \bibinfo{journal}{Nature communications} \textbf{\bibinfo{volume}{4}},
  \bibinfo{pages}{1348} (\bibinfo{year}{2013}).

\bibitem[{\citenamefont{Vaezi}(2013)}]{Vaezi13}
\bibinfo{author}{\bibfnamefont{A.}~\bibnamefont{Vaezi}},
  \bibinfo{journal}{Phys. Rev. B} \textbf{\bibinfo{volume}{87}},
  \bibinfo{pages}{035132} (\bibinfo{year}{2013}),
  \urlprefix\url{https://link.aps.org/doi/10.1103/PhysRevB.87.035132}.

\bibitem[{\citenamefont{Milsted et~al.}(2014)\citenamefont{Milsted, Cobanera,
  Burrello, and Ortiz}}]{Milsted14}
\bibinfo{author}{\bibfnamefont{A.}~\bibnamefont{Milsted}},
  \bibinfo{author}{\bibfnamefont{E.}~\bibnamefont{Cobanera}},
  \bibinfo{author}{\bibfnamefont{M.}~\bibnamefont{Burrello}}, \bibnamefont{and}
  \bibinfo{author}{\bibfnamefont{G.}~\bibnamefont{Ortiz}},
  \bibinfo{journal}{Phys. Rev. B} \textbf{\bibinfo{volume}{90}},
  \bibinfo{pages}{195101} (\bibinfo{year}{2014}),
  \urlprefix\url{https://link.aps.org/doi/10.1103/PhysRevB.90.195101}.

\bibitem[{\citenamefont{Klinovaja et~al.}(2014)\citenamefont{Klinovaja, Yacoby,
  and Loss}}]{Klinovaja14}
\bibinfo{author}{\bibfnamefont{J.}~\bibnamefont{Klinovaja}},
  \bibinfo{author}{\bibfnamefont{A.}~\bibnamefont{Yacoby}}, \bibnamefont{and}
  \bibinfo{author}{\bibfnamefont{D.}~\bibnamefont{Loss}},
  \bibinfo{journal}{Phys. Rev. B} \textbf{\bibinfo{volume}{90}},
  \bibinfo{pages}{155447} (\bibinfo{year}{2014}),
  \urlprefix\url{https://link.aps.org/doi/10.1103/PhysRevB.90.155447}.

\bibitem[{\citenamefont{Alicea and Fendley}(2016)}]{Alicea16}
\bibinfo{author}{\bibfnamefont{J.}~\bibnamefont{Alicea}} \bibnamefont{and}
  \bibinfo{author}{\bibfnamefont{P.}~\bibnamefont{Fendley}},
  \bibinfo{journal}{Annual Review of Condensed Matter Physics}
  \textbf{\bibinfo{volume}{7}}, \bibinfo{pages}{119} (\bibinfo{year}{2016}),
  \urlprefix\url{https://doi.org/10.1146/annurev-conmatphys-031115-011336}.

\bibitem[{\citenamefont{Sagi et~al.}(2017)\citenamefont{Sagi, Haim, Berg, von
  Oppen, and Oreg}}]{Sagi17}
\bibinfo{author}{\bibfnamefont{E.}~\bibnamefont{Sagi}},
  \bibinfo{author}{\bibfnamefont{A.}~\bibnamefont{Haim}},
  \bibinfo{author}{\bibfnamefont{E.}~\bibnamefont{Berg}},
  \bibinfo{author}{\bibfnamefont{F.}~\bibnamefont{von Oppen}},
  \bibnamefont{and} \bibinfo{author}{\bibfnamefont{Y.}~\bibnamefont{Oreg}},
  \bibinfo{journal}{Phys. Rev. B} \textbf{\bibinfo{volume}{96}},
  \bibinfo{pages}{235144} (\bibinfo{year}{2017}),
  \urlprefix\url{https://link.aps.org/doi/10.1103/PhysRevB.96.235144}.

\bibitem[{\citenamefont{Liang et~al.}(2019)\citenamefont{Liang, Simion, and
  Lyanda-Geller}}]{Liang19}
\bibinfo{author}{\bibfnamefont{J.}~\bibnamefont{Liang}},
  \bibinfo{author}{\bibfnamefont{G.}~\bibnamefont{Simion}}, \bibnamefont{and}
  \bibinfo{author}{\bibfnamefont{Y.}~\bibnamefont{Lyanda-Geller}},
  \bibinfo{journal}{Phys. Rev. B} \textbf{\bibinfo{volume}{100}},
  \bibinfo{pages}{075155} (\bibinfo{year}{2019}),
  \urlprefix\url{https://link.aps.org/doi/10.1103/PhysRevB.100.075155}.

\bibitem[{\citenamefont{Vaezi}(2014)}]{Vaezi14}
\bibinfo{author}{\bibfnamefont{A.}~\bibnamefont{Vaezi}},
  \bibinfo{journal}{Phys. Rev. X} \textbf{\bibinfo{volume}{4}},
  \bibinfo{pages}{031009} (\bibinfo{year}{2014}),
  \urlprefix\url{https://link.aps.org/doi/10.1103/PhysRevX.4.031009}.

\bibitem[{\citenamefont{Mong et~al.}(2017)\citenamefont{Mong, Zaletel,
  Pollmann, and Papi\ifmmode~\acute{c}\else \'{c}\fi{}}}]{Mong15}
\bibinfo{author}{\bibfnamefont{R.~S.~K.} \bibnamefont{Mong}},
  \bibinfo{author}{\bibfnamefont{M.~P.} \bibnamefont{Zaletel}},
  \bibinfo{author}{\bibfnamefont{F.}~\bibnamefont{Pollmann}}, \bibnamefont{and}
  \bibinfo{author}{\bibfnamefont{Z.}~\bibnamefont{Papi\ifmmode~\acute{c}\else
  \'{c}\fi{}}}, \bibinfo{journal}{Phys. Rev. B} \textbf{\bibinfo{volume}{95}},
  \bibinfo{pages}{115136} (\bibinfo{year}{2017}),
  \urlprefix\url{http://link.aps.org/doi/10.1103/PhysRevB.95.115136}.

\bibitem[{\citenamefont{Hu and Kane}(2018)}]{Hu18}
\bibinfo{author}{\bibfnamefont{Y.}~\bibnamefont{Hu}} \bibnamefont{and}
  \bibinfo{author}{\bibfnamefont{C.~L.} \bibnamefont{Kane}},
  \bibinfo{journal}{Phys. Rev. Lett.} \textbf{\bibinfo{volume}{120}},
  \bibinfo{pages}{066801} (\bibinfo{year}{2018}),
  \urlprefix\url{https://link.aps.org/doi/10.1103/PhysRevLett.120.066801}.

\bibitem[{\citenamefont{Lopes et~al.}(2019)\citenamefont{Lopes, Quito, Han, and
  Teo}}]{Lopes19}
\bibinfo{author}{\bibfnamefont{P.~L.~S.} \bibnamefont{Lopes}},
  \bibinfo{author}{\bibfnamefont{V.~L.} \bibnamefont{Quito}},
  \bibinfo{author}{\bibfnamefont{B.}~\bibnamefont{Han}}, \bibnamefont{and}
  \bibinfo{author}{\bibfnamefont{J.~C.~Y.} \bibnamefont{Teo}},
  \bibinfo{journal}{Phys. Rev. B} \textbf{\bibinfo{volume}{100}},
  \bibinfo{pages}{085116} (\bibinfo{year}{2019}),
  \urlprefix\url{https://link.aps.org/doi/10.1103/PhysRevB.100.085116}.

\bibitem[{\citenamefont{G\"ul et~al.}(2022)\citenamefont{G\"ul, Ronen, Lee,
  Shapourian, Zauberman, Lee, Watanabe, Taniguchi, Vishwanath, Yacoby
  et~al.}}]{Gul22}
\bibinfo{author}{\bibfnamefont{O.}~\bibnamefont{G\"ul}},
  \bibinfo{author}{\bibfnamefont{Y.}~\bibnamefont{Ronen}},
  \bibinfo{author}{\bibfnamefont{S.~Y.} \bibnamefont{Lee}},
  \bibinfo{author}{\bibfnamefont{H.}~\bibnamefont{Shapourian}},
  \bibinfo{author}{\bibfnamefont{J.}~\bibnamefont{Zauberman}},
  \bibinfo{author}{\bibfnamefont{Y.~H.} \bibnamefont{Lee}},
  \bibinfo{author}{\bibfnamefont{K.}~\bibnamefont{Watanabe}},
  \bibinfo{author}{\bibfnamefont{T.}~\bibnamefont{Taniguchi}},
  \bibinfo{author}{\bibfnamefont{A.}~\bibnamefont{Vishwanath}},
  \bibinfo{author}{\bibfnamefont{A.}~\bibnamefont{Yacoby}},
  \bibnamefont{et~al.}, \bibinfo{journal}{Phys. Rev. X}
  \textbf{\bibinfo{volume}{12}}, \bibinfo{pages}{021057}
  (\bibinfo{year}{2022}),
  \urlprefix\url{https://link.aps.org/doi/10.1103/PhysRevX.12.021057}.

\bibitem[{\citenamefont{Te\ifmmode \check{s}\else
  \v{s}\fi{}anovi\ifmmode~\acute{c}\else \'{c}\fi{}
  et~al.}(1989)\citenamefont{Te\ifmmode \check{s}\else
  \v{s}\fi{}anovi\ifmmode~\acute{c}\else \'{c}\fi{}, Rasolt, and
  Xing}}]{Teifmmode89}
\bibinfo{author}{\bibfnamefont{Z.}~\bibnamefont{Te\ifmmode \check{s}\else
  \v{s}\fi{}anovi\ifmmode~\acute{c}\else \'{c}\fi{}}},
  \bibinfo{author}{\bibfnamefont{M.}~\bibnamefont{Rasolt}}, \bibnamefont{and}
  \bibinfo{author}{\bibfnamefont{L.}~\bibnamefont{Xing}},
  \bibinfo{journal}{Phys. Rev. Lett.} \textbf{\bibinfo{volume}{63}},
  \bibinfo{pages}{2425} (\bibinfo{year}{1989}),
  \urlprefix\url{https://link.aps.org/doi/10.1103/PhysRevLett.63.2425}.

\bibitem[{\citenamefont{Rajagopal and Vasudevan}(1991)}]{Rajagopal91}
\bibinfo{author}{\bibfnamefont{A.~K.} \bibnamefont{Rajagopal}}
  \bibnamefont{and}
  \bibinfo{author}{\bibfnamefont{R.}~\bibnamefont{Vasudevan}},
  \bibinfo{journal}{Phys. Rev. B} \textbf{\bibinfo{volume}{44}},
  \bibinfo{pages}{2807} (\bibinfo{year}{1991}),
  \urlprefix\url{https://link.aps.org/doi/10.1103/PhysRevB.44.2807}.

\bibitem[{\citenamefont{Te\ifmmode \check{s}\else
  \v{s}\fi{}anovi\ifmmode~\acute{c}\else \'{c}\fi{}
  et~al.}(1991)\citenamefont{Te\ifmmode \check{s}\else
  \v{s}\fi{}anovi\ifmmode~\acute{c}\else \'{c}\fi{}, Rasolt, and
  Xing}}]{Teifmmode91}
\bibinfo{author}{\bibfnamefont{Z.}~\bibnamefont{Te\ifmmode \check{s}\else
  \v{s}\fi{}anovi\ifmmode~\acute{c}\else \'{c}\fi{}}},
  \bibinfo{author}{\bibfnamefont{M.}~\bibnamefont{Rasolt}}, \bibnamefont{and}
  \bibinfo{author}{\bibfnamefont{L.}~\bibnamefont{Xing}},
  \bibinfo{journal}{Phys. Rev. B} \textbf{\bibinfo{volume}{43}},
  \bibinfo{pages}{288} (\bibinfo{year}{1991}),
  \urlprefix\url{https://link.aps.org/doi/10.1103/PhysRevB.43.288}.

\bibitem[{\citenamefont{Norman}(1991)}]{Norman91}
\bibinfo{author}{\bibfnamefont{M.~R.} \bibnamefont{Norman}},
  \bibinfo{journal}{Phys. Rev. Lett.} \textbf{\bibinfo{volume}{66}},
  \bibinfo{pages}{842} (\bibinfo{year}{1991}),
  \urlprefix\url{https://link.aps.org/doi/10.1103/PhysRevLett.66.842}.

\bibitem[{\citenamefont{Akera et~al.}(1991)\citenamefont{Akera, MacDonald,
  Girvin, and Norman}}]{Akera91}
\bibinfo{author}{\bibfnamefont{H.}~\bibnamefont{Akera}},
  \bibinfo{author}{\bibfnamefont{A.~H.} \bibnamefont{MacDonald}},
  \bibinfo{author}{\bibfnamefont{S.~M.} \bibnamefont{Girvin}},
  \bibnamefont{and} \bibinfo{author}{\bibfnamefont{M.~R.}
  \bibnamefont{Norman}}, \bibinfo{journal}{Phys. Rev. Lett.}
  \textbf{\bibinfo{volume}{67}}, \bibinfo{pages}{2375} (\bibinfo{year}{1991}),
  \urlprefix\url{https://link.aps.org/doi/10.1103/PhysRevLett.67.2375}.

\bibitem[{\citenamefont{Rajagopal and Ryan}(1991)}]{Rajagopal91b}
\bibinfo{author}{\bibfnamefont{A.~K.} \bibnamefont{Rajagopal}}
  \bibnamefont{and} \bibinfo{author}{\bibfnamefont{J.~C.} \bibnamefont{Ryan}},
  \bibinfo{journal}{Phys. Rev. B} \textbf{\bibinfo{volume}{44}},
  \bibinfo{pages}{10280} (\bibinfo{year}{1991}),
  \urlprefix\url{https://link.aps.org/doi/10.1103/PhysRevB.44.10280}.

\bibitem[{\citenamefont{Rasolt and Tes\ifmmode \breve{}\else
  \u{}\fi{}anovi\ifmmode~\acute{c}\else \'{c}\fi{}}(1992)}]{Rasolt92}
\bibinfo{author}{\bibfnamefont{M.}~\bibnamefont{Rasolt}} \bibnamefont{and}
  \bibinfo{author}{\bibfnamefont{Z.}~\bibnamefont{Tes\ifmmode \breve{}\else
  \u{}\fi{}anovi\ifmmode~\acute{c}\else \'{c}\fi{}}}, \bibinfo{journal}{Rev.
  Mod. Phys.} \textbf{\bibinfo{volume}{64}}, \bibinfo{pages}{709}
  (\bibinfo{year}{1992}),
  \urlprefix\url{https://link.aps.org/doi/10.1103/RevModPhys.64.709}.

\bibitem[{\citenamefont{MacDonald et~al.}(1992)\citenamefont{MacDonald, Akera,
  and Norman}}]{MacDonald92}
\bibinfo{author}{\bibfnamefont{A.~H.} \bibnamefont{MacDonald}},
  \bibinfo{author}{\bibfnamefont{H.}~\bibnamefont{Akera}}, \bibnamefont{and}
  \bibinfo{author}{\bibfnamefont{M.~R.} \bibnamefont{Norman}},
  \bibinfo{journal}{Phys. Rev. B} \textbf{\bibinfo{volume}{45}},
  \bibinfo{pages}{10147} (\bibinfo{year}{1992}),
  \urlprefix\url{https://link.aps.org/doi/10.1103/PhysRevB.45.10147}.

\bibitem[{\citenamefont{Norman et~al.}(1992)\citenamefont{Norman, Akera, and
  MacDonald}}]{Norman92}
\bibinfo{author}{\bibfnamefont{M.}~\bibnamefont{Norman}},
  \bibinfo{author}{\bibfnamefont{H.}~\bibnamefont{Akera}}, \bibnamefont{and}
  \bibinfo{author}{\bibfnamefont{A.}~\bibnamefont{MacDonald}},
  \bibinfo{journal}{Physica C: Superconductivity}
  \textbf{\bibinfo{volume}{196}}, \bibinfo{pages}{43} (\bibinfo{year}{1992}),
  ISSN \bibinfo{issn}{0921-4534},
  \urlprefix\url{https://www.sciencedirect.com/science/article/pii/092145349290135Y}.

\bibitem[{\citenamefont{Rajagopal}(1992)}]{Rajagopal92}
\bibinfo{author}{\bibfnamefont{A.~K.} \bibnamefont{Rajagopal}},
  \bibinfo{journal}{Phys. Rev. B} \textbf{\bibinfo{volume}{46}},
  \bibinfo{pages}{1224} (\bibinfo{year}{1992}),
  \urlprefix\url{https://link.aps.org/doi/10.1103/PhysRevB.46.1224}.

\bibitem[{\citenamefont{MacDonald et~al.}(1993)\citenamefont{MacDonald, Akera,
  and Norman}}]{MacDonald93}
\bibinfo{author}{\bibfnamefont{A.}~\bibnamefont{MacDonald}},
  \bibinfo{author}{\bibfnamefont{H.}~\bibnamefont{Akera}}, \bibnamefont{and}
  \bibinfo{author}{\bibfnamefont{M.}~\bibnamefont{Norman}},
  \bibinfo{journal}{Australian Journal of Physics}
  \textbf{\bibinfo{volume}{46}}, \bibinfo{pages}{333} (\bibinfo{year}{1993}),
  \urlprefix\url{https://doi.org/10.1071/PH930333}.

\bibitem[{\citenamefont{Ryan and Rajagopal}(1993{\natexlab{a}})}]{Ryan93}
\bibinfo{author}{\bibfnamefont{J.}~\bibnamefont{Ryan}} \bibnamefont{and}
  \bibinfo{author}{\bibfnamefont{A.}~\bibnamefont{Rajagopal}},
  \bibinfo{journal}{Journal of Physics and Chemistry of Solids}
  \textbf{\bibinfo{volume}{54}}, \bibinfo{pages}{1281}
  (\bibinfo{year}{1993}{\natexlab{a}}), ISSN \bibinfo{issn}{0022-3697},
  \bibinfo{note}{special Issue Spectroscopies in Novel Superconductors},
  \urlprefix\url{https://www.sciencedirect.com/science/article/pii/002236979390180Y}.

\bibitem[{\citenamefont{Ryan and Rajagopal}(1993{\natexlab{b}})}]{Ryan93b}
\bibinfo{author}{\bibfnamefont{J.~C.} \bibnamefont{Ryan}} \bibnamefont{and}
  \bibinfo{author}{\bibfnamefont{A.~K.} \bibnamefont{Rajagopal}},
  \bibinfo{journal}{Phys. Rev. B} \textbf{\bibinfo{volume}{47}},
  \bibinfo{pages}{8843} (\bibinfo{year}{1993}{\natexlab{b}}),
  \urlprefix\url{https://link.aps.org/doi/10.1103/PhysRevB.47.8843}.

\bibitem[{\citenamefont{Norman et~al.}(1995)\citenamefont{Norman, MacDonald,
  and Akera}}]{Norman95}
\bibinfo{author}{\bibfnamefont{M.~R.} \bibnamefont{Norman}},
  \bibinfo{author}{\bibfnamefont{A.~H.} \bibnamefont{MacDonald}},
  \bibnamefont{and} \bibinfo{author}{\bibfnamefont{H.}~\bibnamefont{Akera}},
  \bibinfo{journal}{Phys. Rev. B} \textbf{\bibinfo{volume}{51}},
  \bibinfo{pages}{5927} (\bibinfo{year}{1995}),
  \urlprefix\url{https://link.aps.org/doi/10.1103/PhysRevB.51.5927}.

\bibitem[{\citenamefont{Ma\ifmmode~\acute{s}\else
  \'{s}\fi{}ka}(2002)}]{Maifmmode02}
\bibinfo{author}{\bibfnamefont{M.~M.} \bibnamefont{Ma\ifmmode~\acute{s}\else
  \'{s}\fi{}ka}}, \bibinfo{journal}{Phys. Rev. B}
  \textbf{\bibinfo{volume}{66}}, \bibinfo{pages}{054533}
  (\bibinfo{year}{2002}),
  \urlprefix\url{https://link.aps.org/doi/10.1103/PhysRevB.66.054533}.

\bibitem[{\citenamefont{Scherpelz et~al.}(2013)\citenamefont{Scherpelz, Wulin,
  \ifmmode~\check{S}\else \v{S}\fi{}op\'{\i}k, Levin, and
  Rajagopal}}]{Scherpelz13}
\bibinfo{author}{\bibfnamefont{P.}~\bibnamefont{Scherpelz}},
  \bibinfo{author}{\bibfnamefont{D.}~\bibnamefont{Wulin}},
  \bibinfo{author}{\bibfnamefont{B.~c.~v.} \bibnamefont{\ifmmode~\check{S}\else
  \v{S}\fi{}op\'{\i}k}},
  \bibinfo{author}{\bibfnamefont{K.}~\bibnamefont{Levin}}, \bibnamefont{and}
  \bibinfo{author}{\bibfnamefont{A.~K.} \bibnamefont{Rajagopal}},
  \bibinfo{journal}{Phys. Rev. B} \textbf{\bibinfo{volume}{87}},
  \bibinfo{pages}{024516} (\bibinfo{year}{2013}),
  \urlprefix\url{https://link.aps.org/doi/10.1103/PhysRevB.87.024516}.

\bibitem[{\citenamefont{Ran et~al.}(2019)\citenamefont{Ran, Liu, Eo, Campbell,
  Neves, Fuhrman, Saha, Eckberg, Kim, Graf et~al.}}]{Ran19}
\bibinfo{author}{\bibfnamefont{S.}~\bibnamefont{Ran}},
  \bibinfo{author}{\bibfnamefont{I.-L.} \bibnamefont{Liu}},
  \bibinfo{author}{\bibfnamefont{Y.~S.} \bibnamefont{Eo}},
  \bibinfo{author}{\bibfnamefont{D.~J.} \bibnamefont{Campbell}},
  \bibinfo{author}{\bibfnamefont{P.~M.} \bibnamefont{Neves}},
  \bibinfo{author}{\bibfnamefont{W.~T.} \bibnamefont{Fuhrman}},
  \bibinfo{author}{\bibfnamefont{S.~R.} \bibnamefont{Saha}},
  \bibinfo{author}{\bibfnamefont{C.}~\bibnamefont{Eckberg}},
  \bibinfo{author}{\bibfnamefont{H.}~\bibnamefont{Kim}},
  \bibinfo{author}{\bibfnamefont{D.}~\bibnamefont{Graf}}, \bibnamefont{et~al.},
  \bibinfo{journal}{Nature Physics} \textbf{\bibinfo{volume}{15}},
  \bibinfo{pages}{1250} (\bibinfo{year}{2019}),
  \urlprefix\url{https://doi.org/10.1038/s41567-019-0670-x}.

\bibitem[{\citenamefont{Kim et~al.}(2019)\citenamefont{Kim, Balram, Taniguchi,
  Watanabe, Jain, and Smet}}]{Kim19}
\bibinfo{author}{\bibfnamefont{Y.}~\bibnamefont{Kim}},
  \bibinfo{author}{\bibfnamefont{A.~C.} \bibnamefont{Balram}},
  \bibinfo{author}{\bibfnamefont{T.}~\bibnamefont{Taniguchi}},
  \bibinfo{author}{\bibfnamefont{K.}~\bibnamefont{Watanabe}},
  \bibinfo{author}{\bibfnamefont{J.~K.} \bibnamefont{Jain}}, \bibnamefont{and}
  \bibinfo{author}{\bibfnamefont{J.~H.} \bibnamefont{Smet}},
  \bibinfo{journal}{Nature Physics} \textbf{\bibinfo{volume}{15}},
  \bibinfo{pages}{154} (\bibinfo{year}{2019}), ISSN \bibinfo{issn}{1745-2481},
  \urlprefix\url{https://doi.org/10.1038/s41567-018-0355-x}.

\bibitem[{\citenamefont{Chaudhary et~al.}(2021)\citenamefont{Chaudhary,
  MacDonald, and Norman}}]{Chaudhary21}
\bibinfo{author}{\bibfnamefont{G.}~\bibnamefont{Chaudhary}},
  \bibinfo{author}{\bibfnamefont{A.~H.} \bibnamefont{MacDonald}},
  \bibnamefont{and} \bibinfo{author}{\bibfnamefont{M.~R.}
  \bibnamefont{Norman}}, \bibinfo{journal}{Phys. Rev. Res.}
  \textbf{\bibinfo{volume}{3}}, \bibinfo{pages}{033260} (\bibinfo{year}{2021}),
  \urlprefix\url{https://link.aps.org/doi/10.1103/PhysRevResearch.3.033260}.

\bibitem[{\citenamefont{Schirmer
  et~al.}(2022{\natexlab{a}})\citenamefont{Schirmer, Liu, and
  Jain}}]{Schirmer22}
\bibinfo{author}{\bibfnamefont{J.}~\bibnamefont{Schirmer}},
  \bibinfo{author}{\bibfnamefont{C.-X.} \bibnamefont{Liu}}, \bibnamefont{and}
  \bibinfo{author}{\bibfnamefont{J.~K.} \bibnamefont{Jain}},
  \bibinfo{journal}{Proceedings of the National Academy of Sciences}
  \textbf{\bibinfo{volume}{119}}, \bibinfo{pages}{e2202948119}
  (\bibinfo{year}{2022}{\natexlab{a}}),
  \eprint{https://www.pnas.org/doi/pdf/10.1073/pnas.2202948119},
  \urlprefix\url{https://www.pnas.org/doi/abs/10.1073/pnas.2202948119}.

\bibitem[{\citenamefont{Schirmer
  et~al.}(2022{\natexlab{b}})\citenamefont{Schirmer, Jain, and
  Liu}}]{Schirmer22a}
\bibinfo{author}{\bibfnamefont{J.}~\bibnamefont{Schirmer}},
  \bibinfo{author}{\bibfnamefont{J.~K.} \bibnamefont{Jain}}, \bibnamefont{and}
  \bibinfo{author}{\bibfnamefont{C.~X.} \bibnamefont{Liu}}
  (\bibinfo{year}{2022}{\natexlab{b}}),
  \urlprefix\url{https://arxiv.org/abs/2211.15001}.

\bibitem[{\citenamefont{Wen}(1990{\natexlab{a}})}]{Wen90a}
\bibinfo{author}{\bibfnamefont{X.~G.} \bibnamefont{Wen}},
  \bibinfo{journal}{Phys. Rev. Lett.} \textbf{\bibinfo{volume}{64}},
  \bibinfo{pages}{2206} (\bibinfo{year}{1990}{\natexlab{a}}),
  \urlprefix\url{http://link.aps.org/doi/10.1103/PhysRevLett.64.2206}.

\bibitem[{\citenamefont{Hatsugai et~al.}(1993)\citenamefont{Hatsugai, Bares,
  and Wen}}]{Hatsugai93}
\bibinfo{author}{\bibfnamefont{Y.}~\bibnamefont{Hatsugai}},
  \bibinfo{author}{\bibfnamefont{P.-A.} \bibnamefont{Bares}}, \bibnamefont{and}
  \bibinfo{author}{\bibfnamefont{X.~G.} \bibnamefont{Wen}},
  \bibinfo{journal}{Phys. Rev. Lett.} \textbf{\bibinfo{volume}{71}},
  \bibinfo{pages}{424} (\bibinfo{year}{1993}),
  \urlprefix\url{http://link.aps.org/doi/10.1103/PhysRevLett.71.424}.

\bibitem[{\citenamefont{Birman}(1969)}]{Birman69}
\bibinfo{author}{\bibfnamefont{J.~S.} \bibnamefont{Birman}},
  \bibinfo{journal}{Communications on Pure and Applied Mathematics}
  \textbf{\bibinfo{volume}{22}}, \bibinfo{pages}{41} (\bibinfo{year}{1969}),
  \eprint{https://onlinelibrary.wiley.com/doi/pdf/10.1002/cpa.3160220104},
  \urlprefix\url{https://onlinelibrary.wiley.com/doi/abs/10.1002/cpa.3160220104}.

\bibitem[{\citenamefont{Einarsson}(1990)}]{Einarsson90}
\bibinfo{author}{\bibfnamefont{T.}~\bibnamefont{Einarsson}},
  \bibinfo{journal}{Phys. Rev. Lett.} \textbf{\bibinfo{volume}{64}},
  \bibinfo{pages}{1995} (\bibinfo{year}{1990}),
  \urlprefix\url{https://link.aps.org/doi/10.1103/PhysRevLett.64.1995}.

\bibitem[{\citenamefont{Wen}(1990{\natexlab{b}})}]{Wen90}
\bibinfo{author}{\bibfnamefont{X.~G.} \bibnamefont{Wen}},
  \bibinfo{journal}{Phys. Rev. B} \textbf{\bibinfo{volume}{41}},
  \bibinfo{pages}{12838} (\bibinfo{year}{1990}{\natexlab{b}}),
  \urlprefix\url{http://link.aps.org/doi/10.1103/PhysRevB.41.12838}.

\bibitem[{\citenamefont{Oshikawa and Senthil}(2006)}]{Oshikawa06}
\bibinfo{author}{\bibfnamefont{M.}~\bibnamefont{Oshikawa}} \bibnamefont{and}
  \bibinfo{author}{\bibfnamefont{T.}~\bibnamefont{Senthil}},
  \bibinfo{journal}{Phys. Rev. Lett.} \textbf{\bibinfo{volume}{96}},
  \bibinfo{pages}{060601} (\bibinfo{year}{2006}),
  \urlprefix\url{https://link.aps.org/doi/10.1103/PhysRevLett.96.060601}.

\bibitem[{\citenamefont{Sato et~al.}(2006)\citenamefont{Sato, Kohmoto, and
  Wu}}]{Sato06}
\bibinfo{author}{\bibfnamefont{M.}~\bibnamefont{Sato}},
  \bibinfo{author}{\bibfnamefont{M.}~\bibnamefont{Kohmoto}}, \bibnamefont{and}
  \bibinfo{author}{\bibfnamefont{Y.-S.} \bibnamefont{Wu}},
  \bibinfo{journal}{Phys. Rev. Lett.} \textbf{\bibinfo{volume}{97}},
  \bibinfo{pages}{010601} (\bibinfo{year}{2006}),
  \urlprefix\url{https://link.aps.org/doi/10.1103/PhysRevLett.97.010601}.

\bibitem[{\citenamefont{Oshikawa et~al.}(2007)\citenamefont{Oshikawa, Kim,
  Shtengel, Nayak, and Tewari}}]{Oshikawa07}
\bibinfo{author}{\bibfnamefont{M.}~\bibnamefont{Oshikawa}},
  \bibinfo{author}{\bibfnamefont{Y.~B.} \bibnamefont{Kim}},
  \bibinfo{author}{\bibfnamefont{K.}~\bibnamefont{Shtengel}},
  \bibinfo{author}{\bibfnamefont{C.}~\bibnamefont{Nayak}}, \bibnamefont{and}
  \bibinfo{author}{\bibfnamefont{S.}~\bibnamefont{Tewari}},
  \bibinfo{journal}{Annals of Physics} \textbf{\bibinfo{volume}{322}},
  \bibinfo{pages}{1477 } (\bibinfo{year}{2007}), ISSN
  \bibinfo{issn}{0003-4916},
  \urlprefix\url{http://www.sciencedirect.com/science/article/pii/S0003491606001837}.

\bibitem[{\citenamefont{Haldane}(1985)}]{Haldane85b}
\bibinfo{author}{\bibfnamefont{F.~D.~M.} \bibnamefont{Haldane}},
  \bibinfo{journal}{Phys. Rev. Lett.} \textbf{\bibinfo{volume}{55}},
  \bibinfo{pages}{2095} (\bibinfo{year}{1985}),
  \urlprefix\url{http://link.aps.org/doi/10.1103/PhysRevLett.55.2095}.

\bibitem[{\citenamefont{Wen and Niu}(1990)}]{Wen90d}
\bibinfo{author}{\bibfnamefont{X.~G.} \bibnamefont{Wen}} \bibnamefont{and}
  \bibinfo{author}{\bibfnamefont{Q.}~\bibnamefont{Niu}},
  \bibinfo{journal}{Phys. Rev. B} \textbf{\bibinfo{volume}{41}},
  \bibinfo{pages}{9377} (\bibinfo{year}{1990}),
  \urlprefix\url{https://link.aps.org/doi/10.1103/PhysRevB.41.9377}.

\bibitem[{\citenamefont{Haldane}(1983)}]{Haldane83}
\bibinfo{author}{\bibfnamefont{F.~D.~M.} \bibnamefont{Haldane}},
  \bibinfo{journal}{Phys. Rev. Lett.} \textbf{\bibinfo{volume}{51}},
  \bibinfo{pages}{605} (\bibinfo{year}{1983}),
  \urlprefix\url{http://link.aps.org/doi/10.1103/PhysRevLett.51.605}.

\bibitem[{\citenamefont{Rashba}(1960)}]{Rashba60}
\bibinfo{author}{\bibfnamefont{E.}~\bibnamefont{Rashba}},
  \bibinfo{journal}{Sov. Phys.-Solid State} \textbf{\bibinfo{volume}{2}},
  \bibinfo{pages}{1109} (\bibinfo{year}{1960}).

\bibitem[{\citenamefont{Bychkov and Rashba}(1984)}]{Yu84}
\bibinfo{author}{\bibfnamefont{Y.~A.} \bibnamefont{Bychkov}} \bibnamefont{and}
  \bibinfo{author}{\bibfnamefont{E.~I.} \bibnamefont{Rashba}},
  \bibinfo{journal}{Journal of Physics C: Solid State Physics}
  \textbf{\bibinfo{volume}{17}}, \bibinfo{pages}{6039} (\bibinfo{year}{1984}),
  \urlprefix\url{https://dx.doi.org/10.1088/0022-3719/17/33/015}.

\bibitem[{\citenamefont{Schliemann et~al.}(2003)\citenamefont{Schliemann,
  Egues, and Loss}}]{Schliemann03}
\bibinfo{author}{\bibfnamefont{J.}~\bibnamefont{Schliemann}},
  \bibinfo{author}{\bibfnamefont{J.~C.} \bibnamefont{Egues}}, \bibnamefont{and}
  \bibinfo{author}{\bibfnamefont{D.}~\bibnamefont{Loss}},
  \bibinfo{journal}{Phys. Rev. B} \textbf{\bibinfo{volume}{67}},
  \bibinfo{pages}{085302} (\bibinfo{year}{2003}),
  \urlprefix\url{https://link.aps.org/doi/10.1103/PhysRevB.67.085302}.

\bibitem[{\citenamefont{Shen et~al.}(2004)\citenamefont{Shen, Ma, Xie, and
  Zhang}}]{Shen04}
\bibinfo{author}{\bibfnamefont{S.-Q.} \bibnamefont{Shen}},
  \bibinfo{author}{\bibfnamefont{M.}~\bibnamefont{Ma}},
  \bibinfo{author}{\bibfnamefont{X.~C.} \bibnamefont{Xie}}, \bibnamefont{and}
  \bibinfo{author}{\bibfnamefont{F.~C.} \bibnamefont{Zhang}},
  \bibinfo{journal}{Phys. Rev. Lett.} \textbf{\bibinfo{volume}{92}},
  \bibinfo{pages}{256603} (\bibinfo{year}{2004}),
  \urlprefix\url{https://link.aps.org/doi/10.1103/PhysRevLett.92.256603}.

\bibitem[{\citenamefont{Ito et~al.}(2012)\citenamefont{Ito, Nomura, and
  Shibata}}]{Ito12}
\bibinfo{author}{\bibfnamefont{T.}~\bibnamefont{Ito}},
  \bibinfo{author}{\bibfnamefont{K.}~\bibnamefont{Nomura}}, \bibnamefont{and}
  \bibinfo{author}{\bibfnamefont{N.}~\bibnamefont{Shibata}},
  \bibinfo{journal}{Journal of the Physical Society of Japan}
  \textbf{\bibinfo{volume}{81}}, \bibinfo{pages}{034713}
  (\bibinfo{year}{2012}), \eprint{https://doi.org/10.1143/JPSJ.81.034713},
  \urlprefix\url{https://doi.org/10.1143/JPSJ.81.034713}.

\bibitem[{\citenamefont{Castro~Neto et~al.}(2009)\citenamefont{Castro~Neto,
  Guinea, Peres, Novoselov, and Geim}}]{Neto09}
\bibinfo{author}{\bibfnamefont{A.~H.} \bibnamefont{Castro~Neto}},
  \bibinfo{author}{\bibfnamefont{F.}~\bibnamefont{Guinea}},
  \bibinfo{author}{\bibfnamefont{N.~M.~R.} \bibnamefont{Peres}},
  \bibinfo{author}{\bibfnamefont{K.~S.} \bibnamefont{Novoselov}},
  \bibnamefont{and} \bibinfo{author}{\bibfnamefont{A.~K.} \bibnamefont{Geim}},
  \bibinfo{journal}{Rev. Mod. Phys.} \textbf{\bibinfo{volume}{81}},
  \bibinfo{pages}{109} (\bibinfo{year}{2009}),
  \urlprefix\url{http://link.aps.org/doi/10.1103/RevModPhys.81.109}.

\bibitem[{\citenamefont{Wu and Yang}(1976)}]{Wu76}
\bibinfo{author}{\bibfnamefont{T.~T.} \bibnamefont{Wu}} \bibnamefont{and}
  \bibinfo{author}{\bibfnamefont{C.~N.} \bibnamefont{Yang}},
  \bibinfo{journal}{Nucl. Phys. B} \textbf{\bibinfo{volume}{107}},
  \bibinfo{pages}{365} (\bibinfo{year}{1976}).

\bibitem[{\citenamefont{Wu and Yang}(1977)}]{Wu77}
\bibinfo{author}{\bibfnamefont{T.~T.} \bibnamefont{Wu}} \bibnamefont{and}
  \bibinfo{author}{\bibfnamefont{C.~N.} \bibnamefont{Yang}},
  \bibinfo{journal}{Phys. Rev. D} \textbf{\bibinfo{volume}{16}},
  \bibinfo{pages}{1018} (\bibinfo{year}{1977}),
  \urlprefix\url{http://link.aps.org/doi/10.1103/PhysRevD.16.1018}.

\bibitem[{\citenamefont{Nomura and MacDonald}(2006)}]{Nomura06}
\bibinfo{author}{\bibfnamefont{K.}~\bibnamefont{Nomura}} \bibnamefont{and}
  \bibinfo{author}{\bibfnamefont{A.~H.} \bibnamefont{MacDonald}},
  \bibinfo{journal}{Phys. Rev. Lett.} \textbf{\bibinfo{volume}{96}},
  \bibinfo{pages}{256602} (\bibinfo{year}{2006}),
  \urlprefix\url{http://link.aps.org/doi/10.1103/PhysRevLett.96.256602}.

\bibitem[{\citenamefont{Goerbig et~al.}(2006)\citenamefont{Goerbig, Moessner,
  and Dou\ifmmode~\mbox{\c{c}}\else \c{c}\fi{}ot}}]{Goerbig06}
\bibinfo{author}{\bibfnamefont{M.~O.} \bibnamefont{Goerbig}},
  \bibinfo{author}{\bibfnamefont{R.}~\bibnamefont{Moessner}}, \bibnamefont{and}
  \bibinfo{author}{\bibfnamefont{B.}~\bibnamefont{Dou\ifmmode~\mbox{\c{c}}\else
  \c{c}\fi{}ot}}, \bibinfo{journal}{Phys. Rev. B}
  \textbf{\bibinfo{volume}{74}}, \bibinfo{pages}{161407}
  (\bibinfo{year}{2006}),
  \urlprefix\url{http://link.aps.org/doi/10.1103/PhysRevB.74.161407}.

\bibitem[{\citenamefont{Apalkov and Chakraborty}(2006)}]{Apalkov06}
\bibinfo{author}{\bibfnamefont{V.~M.} \bibnamefont{Apalkov}} \bibnamefont{and}
  \bibinfo{author}{\bibfnamefont{T.}~\bibnamefont{Chakraborty}},
  \bibinfo{journal}{Phys. Rev. Lett.} \textbf{\bibinfo{volume}{97}},
  \bibinfo{pages}{126801} (\bibinfo{year}{2006}),
  \urlprefix\url{http://link.aps.org/doi/10.1103/PhysRevLett.97.126801}.

\bibitem[{\citenamefont{T\ifmmode~\mbox{\H{o}}\else \H{o}\fi{}ke
  et~al.}(2006)\citenamefont{T\ifmmode~\mbox{\H{o}}\else \H{o}\fi{}ke, Lammert,
  Crespi, and Jain}}]{Toke06}
\bibinfo{author}{\bibfnamefont{C.}~\bibnamefont{T\ifmmode~\mbox{\H{o}}\else
  \H{o}\fi{}ke}}, \bibinfo{author}{\bibfnamefont{P.~E.} \bibnamefont{Lammert}},
  \bibinfo{author}{\bibfnamefont{V.~H.} \bibnamefont{Crespi}},
  \bibnamefont{and} \bibinfo{author}{\bibfnamefont{J.~K.} \bibnamefont{Jain}},
  \bibinfo{journal}{Phys. Rev. B} \textbf{\bibinfo{volume}{74}},
  \bibinfo{pages}{235417} (\bibinfo{year}{2006}),
  \urlprefix\url{http://link.aps.org/doi/10.1103/PhysRevB.74.235417}.

\bibitem[{\citenamefont{T\ifmmode~\mbox{\H{o}}\else \H{o}\fi{}ke and
  Jain}(2007)}]{Toke07}
\bibinfo{author}{\bibfnamefont{C.}~\bibnamefont{T\ifmmode~\mbox{\H{o}}\else
  \H{o}\fi{}ke}} \bibnamefont{and} \bibinfo{author}{\bibfnamefont{J.~K.}
  \bibnamefont{Jain}}, \bibinfo{journal}{Phys. Rev. B}
  \textbf{\bibinfo{volume}{75}}, \bibinfo{pages}{245440}
  (\bibinfo{year}{2007}),
  \urlprefix\url{http://link.aps.org/doi/10.1103/PhysRevB.75.245440}.

\bibitem[{\citenamefont{Balram et~al.}(2015)\citenamefont{Balram,
  T\ifmmode~\mbox{\H{o}}\else \H{o}\fi{}ke, W\'ojs, and Jain}}]{Balram15c}
\bibinfo{author}{\bibfnamefont{A.~C.} \bibnamefont{Balram}},
  \bibinfo{author}{\bibfnamefont{C.}~\bibnamefont{T\ifmmode~\mbox{\H{o}}\else
  \H{o}\fi{}ke}}, \bibinfo{author}{\bibfnamefont{A.}~\bibnamefont{W\'ojs}},
  \bibnamefont{and} \bibinfo{author}{\bibfnamefont{J.~K.} \bibnamefont{Jain}},
  \bibinfo{journal}{Phys. Rev. B} \textbf{\bibinfo{volume}{92}},
  \bibinfo{pages}{205120} (\bibinfo{year}{2015}),
  \urlprefix\url{http://link.aps.org/doi/10.1103/PhysRevB.92.205120}.

\bibitem[{\citenamefont{Jellal}(2008)}]{Jellal08}
\bibinfo{author}{\bibfnamefont{A.}~\bibnamefont{Jellal}},
  \bibinfo{journal}{Nucl. Phys. B} \textbf{\bibinfo{volume}{804}},
  \bibinfo{pages}{361 } (\bibinfo{year}{2008}), ISSN \bibinfo{issn}{0550-3213},
  \urlprefix\url{http://www.sciencedirect.com/science/article/pii/S0550321308002307}.

\bibitem[{\citenamefont{Arciniaga and Peterson}(2016)}]{Arciniaga16}
\bibinfo{author}{\bibfnamefont{M.}~\bibnamefont{Arciniaga}} \bibnamefont{and}
  \bibinfo{author}{\bibfnamefont{M.~R.} \bibnamefont{Peterson}},
  \bibinfo{journal}{Phys. Rev. B} \textbf{\bibinfo{volume}{94}},
  \bibinfo{pages}{035105} (\bibinfo{year}{2016}),
  \urlprefix\url{http://link.aps.org/doi/10.1103/PhysRevB.94.035105}.

\bibitem[{\citenamefont{Hasebe}(2016)}]{Hasebe16}
\bibinfo{author}{\bibfnamefont{K.}~\bibnamefont{Hasebe}},
  \bibinfo{journal}{International Journal of Modern Physics A}
  \textbf{\bibinfo{volume}{31}}, \bibinfo{pages}{1650117}
  (\bibinfo{year}{2016}),
  \eprint{http://www.worldscientific.com/doi/pdf/10.1142/S0217751X16501177},
  \urlprefix\url{http://www.worldscientific.com/doi/abs/10.1142/S0217751X16501177}.

\bibitem[{\citenamefont{Yonaga et~al.}(2016)\citenamefont{Yonaga, Hasebe, and
  Shibata}}]{Yonaga16}
\bibinfo{author}{\bibfnamefont{K.}~\bibnamefont{Yonaga}},
  \bibinfo{author}{\bibfnamefont{K.}~\bibnamefont{Hasebe}}, \bibnamefont{and}
  \bibinfo{author}{\bibfnamefont{N.}~\bibnamefont{Shibata}},
  \bibinfo{journal}{Phys. Rev. B} \textbf{\bibinfo{volume}{93}},
  \bibinfo{pages}{235122} (\bibinfo{year}{2016}),
  \urlprefix\url{http://link.aps.org/doi/10.1103/PhysRevB.93.235122}.

\bibitem[{\citenamefont{Dodgson}(1996)}]{Dodgson_96}
\bibinfo{author}{\bibfnamefont{M.~J.~W.} \bibnamefont{Dodgson}},
  \bibinfo{journal}{Journal of Physics A: Mathematical and General}
  \textbf{\bibinfo{volume}{29}}, \bibinfo{pages}{2499} (\bibinfo{year}{1996}),
  \urlprefix\url{https://dx.doi.org/10.1088/0305-4470/29/10/028}.

\bibitem[{\citenamefont{Dodgson and Moore}(1997)}]{Dodgson97}
\bibinfo{author}{\bibfnamefont{M.~J.~W.} \bibnamefont{Dodgson}}
  \bibnamefont{and} \bibinfo{author}{\bibfnamefont{M.~A.} \bibnamefont{Moore}},
  \bibinfo{journal}{Phys. Rev. B} \textbf{\bibinfo{volume}{55}},
  \bibinfo{pages}{3816} (\bibinfo{year}{1997}),
  \urlprefix\url{https://link.aps.org/doi/10.1103/PhysRevB.55.3816}.

\bibitem[{\citenamefont{Kleiner et~al.}(1964)\citenamefont{Kleiner, Roth, and
  Autler}}]{Kleiner64}
\bibinfo{author}{\bibfnamefont{W.~H.} \bibnamefont{Kleiner}},
  \bibinfo{author}{\bibfnamefont{L.~M.} \bibnamefont{Roth}}, \bibnamefont{and}
  \bibinfo{author}{\bibfnamefont{S.~H.} \bibnamefont{Autler}},
  \bibinfo{journal}{Phys. Rev.} \textbf{\bibinfo{volume}{133}},
  \bibinfo{pages}{A1226} (\bibinfo{year}{1964}),
  \urlprefix\url{https://link.aps.org/doi/10.1103/PhysRev.133.A1226}.

\bibitem[{\citenamefont{Thomson}(1904)}]{Thomson04}
\bibinfo{author}{\bibfnamefont{J.~J.} \bibnamefont{Thomson}},
  \bibinfo{journal}{Phil. Mag.} \textbf{\bibinfo{volume}{7}},
  \bibinfo{pages}{237} (\bibinfo{year}{1904}).

\bibitem[{\citenamefont{Wales and Ulker}(2006)}]{Wales06}
\bibinfo{author}{\bibfnamefont{D.~J.} \bibnamefont{Wales}} \bibnamefont{and}
  \bibinfo{author}{\bibfnamefont{S.}~\bibnamefont{Ulker}},
  \bibinfo{journal}{Phys. Rev. B} \textbf{\bibinfo{volume}{74}},
  \bibinfo{pages}{212101} (\bibinfo{year}{2006}),
  \urlprefix\url{https://link.aps.org/doi/10.1103/PhysRevB.74.212101}.

\bibitem[{\citenamefont{Wales et~al.}(2009)\citenamefont{Wales, McKay, and
  Altschuler}}]{Wales09}
\bibinfo{author}{\bibfnamefont{D.~J.} \bibnamefont{Wales}},
  \bibinfo{author}{\bibfnamefont{H.}~\bibnamefont{McKay}}, \bibnamefont{and}
  \bibinfo{author}{\bibfnamefont{E.~L.} \bibnamefont{Altschuler}},
  \bibinfo{journal}{Phys. Rev. B} \textbf{\bibinfo{volume}{79}},
  \bibinfo{pages}{224115} (\bibinfo{year}{2009}),
  \urlprefix\url{https://link.aps.org/doi/10.1103/PhysRevB.79.224115}.

\bibitem[{\citenamefont{Zhao et~al.}(2018)\citenamefont{Zhao, Zhang, and
  Jain}}]{Zhao18}
\bibinfo{author}{\bibfnamefont{J.}~\bibnamefont{Zhao}},
  \bibinfo{author}{\bibfnamefont{Y.}~\bibnamefont{Zhang}}, \bibnamefont{and}
  \bibinfo{author}{\bibfnamefont{J.~K.} \bibnamefont{Jain}},
  \bibinfo{journal}{Phys. Rev. Lett.} \textbf{\bibinfo{volume}{121}},
  \bibinfo{pages}{116802} (\bibinfo{year}{2018}),
  \urlprefix\url{https://link.aps.org/doi/10.1103/PhysRevLett.121.116802}.

\bibitem[{\citenamefont{Ando}(1983)}]{Ando83}
\bibinfo{author}{\bibfnamefont{T.}~\bibnamefont{Ando}},
  \bibinfo{journal}{Journal of the Physical Society of Japan}
  \textbf{\bibinfo{volume}{52}}, \bibinfo{pages}{1740} (\bibinfo{year}{1983}),
  \urlprefix\url{https://doi.org/10.1143/JPSJ.52.1740}.

\bibitem[{\citenamefont{Ryu and Hatsugai}(2006)}]{Ryu06}
\bibinfo{author}{\bibfnamefont{S.}~\bibnamefont{Ryu}} \bibnamefont{and}
  \bibinfo{author}{\bibfnamefont{Y.}~\bibnamefont{Hatsugai}},
  \bibinfo{journal}{Phys. Rev. B} \textbf{\bibinfo{volume}{73}},
  \bibinfo{pages}{245115} (\bibinfo{year}{2006}),
  \urlprefix\url{https://link.aps.org/doi/10.1103/PhysRevB.73.245115}.

\bibitem[{\citenamefont{Li and Haldane}(2008)}]{Li08}
\bibinfo{author}{\bibfnamefont{H.}~\bibnamefont{Li}} \bibnamefont{and}
  \bibinfo{author}{\bibfnamefont{F.~D.~M.} \bibnamefont{Haldane}},
  \bibinfo{journal}{Phys. Rev. Lett.} \textbf{\bibinfo{volume}{101}},
  \bibinfo{pages}{010504} (\bibinfo{year}{2008}),
  \urlprefix\url{http://link.aps.org/doi/10.1103/PhysRevLett.101.010504}.

\bibitem[{\citenamefont{Prodan et~al.}(2010)\citenamefont{Prodan, Hughes, and
  Bernevig}}]{Prodan10}
\bibinfo{author}{\bibfnamefont{E.}~\bibnamefont{Prodan}},
  \bibinfo{author}{\bibfnamefont{T.~L.} \bibnamefont{Hughes}},
  \bibnamefont{and} \bibinfo{author}{\bibfnamefont{B.~A.}
  \bibnamefont{Bernevig}}, \bibinfo{journal}{Phys. Rev. Lett.}
  \textbf{\bibinfo{volume}{105}}, \bibinfo{pages}{115501}
  (\bibinfo{year}{2010}),
  \urlprefix\url{https://link.aps.org/doi/10.1103/PhysRevLett.105.115501}.

\bibitem[{\citenamefont{Peschel}(2003)}]{Peschel03}
\bibinfo{author}{\bibfnamefont{I.}~\bibnamefont{Peschel}},
  \bibinfo{journal}{Journal of Physics A: Mathematical and General}
  \textbf{\bibinfo{volume}{36}}, \bibinfo{pages}{L205} (\bibinfo{year}{2003}),
  \urlprefix\url{https://dx.doi.org/10.1088/0305-4470/36/14/101}.

\bibitem[{\citenamefont{Cheong and Henley}(2004)}]{Cheong04}
\bibinfo{author}{\bibfnamefont{S.-A.} \bibnamefont{Cheong}} \bibnamefont{and}
  \bibinfo{author}{\bibfnamefont{C.~L.} \bibnamefont{Henley}},
  \bibinfo{journal}{Phys. Rev. B} \textbf{\bibinfo{volume}{69}},
  \bibinfo{pages}{075111} (\bibinfo{year}{2004}),
  \urlprefix\url{https://link.aps.org/doi/10.1103/PhysRevB.69.075111}.

\bibitem[{\citenamefont{Oliveira et~al.}(2014)\citenamefont{Oliveira, Ribeiro,
  and Sacramento}}]{Oliveira14}
\bibinfo{author}{\bibfnamefont{T.~P.} \bibnamefont{Oliveira}},
  \bibinfo{author}{\bibfnamefont{P.}~\bibnamefont{Ribeiro}}, \bibnamefont{and}
  \bibinfo{author}{\bibfnamefont{P.~D.} \bibnamefont{Sacramento}},
  \bibinfo{journal}{Journal of Physics: Condensed Matter}
  \textbf{\bibinfo{volume}{26}}, \bibinfo{pages}{425702}
  (\bibinfo{year}{2014}),
  \urlprefix\url{https://dx.doi.org/10.1088/0953-8984/26/42/425702}.

\bibitem[{\citenamefont{Rodr\'{\i}guez and Sierra}(2009)}]{Rodriguez09}
\bibinfo{author}{\bibfnamefont{I.~D.} \bibnamefont{Rodr\'{\i}guez}}
  \bibnamefont{and} \bibinfo{author}{\bibfnamefont{G.}~\bibnamefont{Sierra}},
  \bibinfo{journal}{Phys. Rev. B} \textbf{\bibinfo{volume}{80}},
  \bibinfo{pages}{153303} (\bibinfo{year}{2009}),
  \urlprefix\url{https://link.aps.org/doi/10.1103/PhysRevB.80.153303}.

\end{thebibliography}
\bibliographystyle{apsrev}

\end{document}